\documentclass[aps,prb,twocolumn,superscriptaddress,longbibliographys]{revtex4-2}
\usepackage{graphicx}
\usepackage{bm}% bold math
\usepackage{color}
\usepackage{epstopdf}
\usepackage{amsmath}
\usepackage{url}
\definecolor{darkGreen}{RGB}{0,110,0}
\definecolor{darkBlue}{RGB}{0,0,130}
\usepackage[colorlinks=true, urlcolor=darkBlue,citecolor=darkGreen,linkcolor=darkBlue]{hyperref}
\begin{document}

%Title of paper
\title{Hallmarks of non-trivial topology in Josephson junctions based on oxide nanochannels}

\author{Alfonso Maiellaro}
\thanks{These two authors contributed equally to this work.}
\affiliation{Dipartimento di Fisica "E.R. Caianiello", Università di Salerno, Via Giovanni Paolo II, 132, I-84084 Fisciano (SA), Italy}
\author{Jacopo Settino}
\thanks{These two authors contributed equally to this work.}
\affiliation{Dipartimento di Fisica, Università della Calabria, Via P. Bucci Arcavacata di Rende (CS), Italy}
\affiliation{INFN, Gruppo collegato di Cosenza, Italy}
\author{Claudio Guarcello}
\affiliation{Dipartimento di Fisica "E.R. Caianiello", Università di Salerno, Via Giovanni Paolo II, 132, I-84084 Fisciano (SA), Italy}
\affiliation{INFN, Sezione di Napoli, Gruppo collegato di Salerno, Italy}
\author{Francesco Romeo}
\affiliation{Dipartimento di Fisica "E.R. Caianiello", Università di Salerno, Via Giovanni Paolo II, 132, I-84084 Fisciano (SA), Italy}
\affiliation{INFN, Sezione di Napoli, Gruppo collegato di Salerno, Italy}
\author{Roberta Citro}
\affiliation{Dipartimento di Fisica "E.R. Caianiello", Università di Salerno, Via Giovanni Paolo II, 132, I-84084 Fisciano (SA), Italy}
\affiliation{INFN, Sezione di Napoli, Gruppo collegato di Salerno, Italy}
\date{\today}

\begin{abstract}
We investigate the topological properties of a Josephson junction obtained by constraining a two-dimensional electron gas at oxide interface to form a quasi-1D conductor. We reveal an anomalous critical current behaviour with a magnetic field applied perpendicular to the Rashba spin-orbit one. We relate the observed critical current enhancement at small magnetic fields with a non-trivial topology, accompanied by Majorana bound states (MBSs) pinned at the edges of the superconducting leads. Signatures of MBSs also include a sawtooth profile in the current-phase relation. Our findings allow to recognize fingerprints of topological superconductivity in non-centrosymmetric materials and confined systems with Rashba spin-orbit interaction, and to explain recent experimental observations for which a microscopic description is still lacking.\\
\end{abstract}
\maketitle

\section{Introduction}
Topological superconductivity (TSC) is an exotic phase of matter in which the fully gapped superconducting bulk hosts Majorana surface states protected by non-Abelian statistics and/or symmetries. In condensed matter systems, the realization of TSC requires the simultaneous presence of an $s$-wave superconducting order (SC) accompanied by the breaking of inversion and time-reversal symmetry. Inversion symmetry breaking might occur in non-centrosymmetric materials and/or confined systems exhibiting large Rashba spin-orbit coupling (SOC). Time-reversal symmetry breaking can be induced by a magnetic order of intrinsic origin or triggered by external magnetic fields. So far, research on TSC has been mainly focused on platforms realized by semiconducting nanowires proximized by a conventional superconductor \cite{Lutchyn_2010,Maiellaro2019,Oreg_2010,sau_2010,Aguado_2017, doi:10.1126/science.1222360,Nat10.1038,Scirep10.1038,Nat17162,zhang_2018}, although other materials and platforms have been recently proposed \cite{condmat6020015,zhang_iron_2018,MaiellaroEPJst,wray_2011,fornieri_2019,condmat7010026}.

Several theoretical proposals suggested that the two-dimensional electron gases (2DEGs) formed at the interface between transition metal oxides, like LaAlO$_3$ and SrTiO$_3$ (LAO/STO) \cite{barthelemy_2021,PhysRevB.103.235120}, are promising candidates for the realization of topological quantum gates, both in 2D \cite{schmalian_2015} and in quasi-one-dimensional (1D) models \cite{mazziotti_2018,PhysRevB.100.094526}. These ideas are based on the extraordinary properties of these materials and, in particular, the simultaneous presence of strong SOC \cite{caviglia_2010} and 2D-SC \cite{reyren_2007}, both electrically tunable \cite{caviglia_2008}. All these phenomena are related to interfacial orbital degrees of freedom, which dominate the oxide 2DEGs physics \cite{salluzzo_2009}.

The evidence of TSC in oxide 2DEGs has remained elusive up to now. Suggestions in this direction come from the anomalous Josephson current magnetic-field dependence in nanobridges at LAO/STO interfaces \cite{PhysRevB.95.140502,NPJ066}. One of the most intriguing result is an anomalous enhancement of the critical current pattern, $I_c(H)$, as a function of a magnetic field $H$ orthogonal to the 2DEG plane. An asymmetry with respect to the direction of the applied magnetic field, $I_c(H)\neq I_c(-H)$, is also reported. The anomalous pattern was interpreted as a possible signature of an unconventional component of the order parameter, giving rise to three channels current with intrinsic phase shift \cite{NPJ066}.\\
On the other hand, a symmetric critical current pattern with a strong enhancement at small magnetic fields has also been observed in a Josephson junction formed by InAs nanowire proximized by Ti/Al superconducting leads. In the latter case, it was speculated that the observed critical supercurrent increase is compatible with a magnetic field-induced topological transition. Despite the relevance of these experimental findings, in both cases a multiband microscopic model demonstrating the connection between the anomalous features of the Josephson current and the topological phase transition is still missing.

Motivated by these experimental observations, we study the transport properties of an oxide-based Josephson junction, made by constraining the 2DEG at LAO/STO (001) interface \cite{PhysRevB.97.174522}, to form a quasi-1D system. Our main result is that the strong enhancement of the critical current with applied magnetic field can be associated with the appearance of MBSs at the edges of the superconducting leads. To argue our conclusions, we match the transport properties of the junction with a microscopic spectral analysis of the system. We find that the maximum of the critical current pattern at finite magnetic field corresponds to a gap closing in the energy spectrum with the appearance of MBSs, as signalled by the Majorana polarization analysis. At the same time, the current-phase relations (CPRs) show a sawtooth profile.  In the absence of MBSs, we observe sinusoidal CPRs and a lowering of the critical current with increasing magnetic fields. Our findings demonstrate that the experimental evidences of anomalous Josephson pattern discussed in \cite{NatCommun14984,NPJ066} are strong hallmarks of topology.

\section{Theoretical model} 
We consider a short superconductor-normal-superconductor (SNS) junction realized by constraining the 2DEG at LAO/STO (001) interface, to form a quasi-1D system. The latter condition is experimentally realizable via appropriate gating of the system. In the absence of superconductivity, the system is described by the nanochannel Hamiltonian 
\begin{figure*}
	\begin{center}
		%\hspace*{1cm}
		\includegraphics[width=0.9\textwidth]{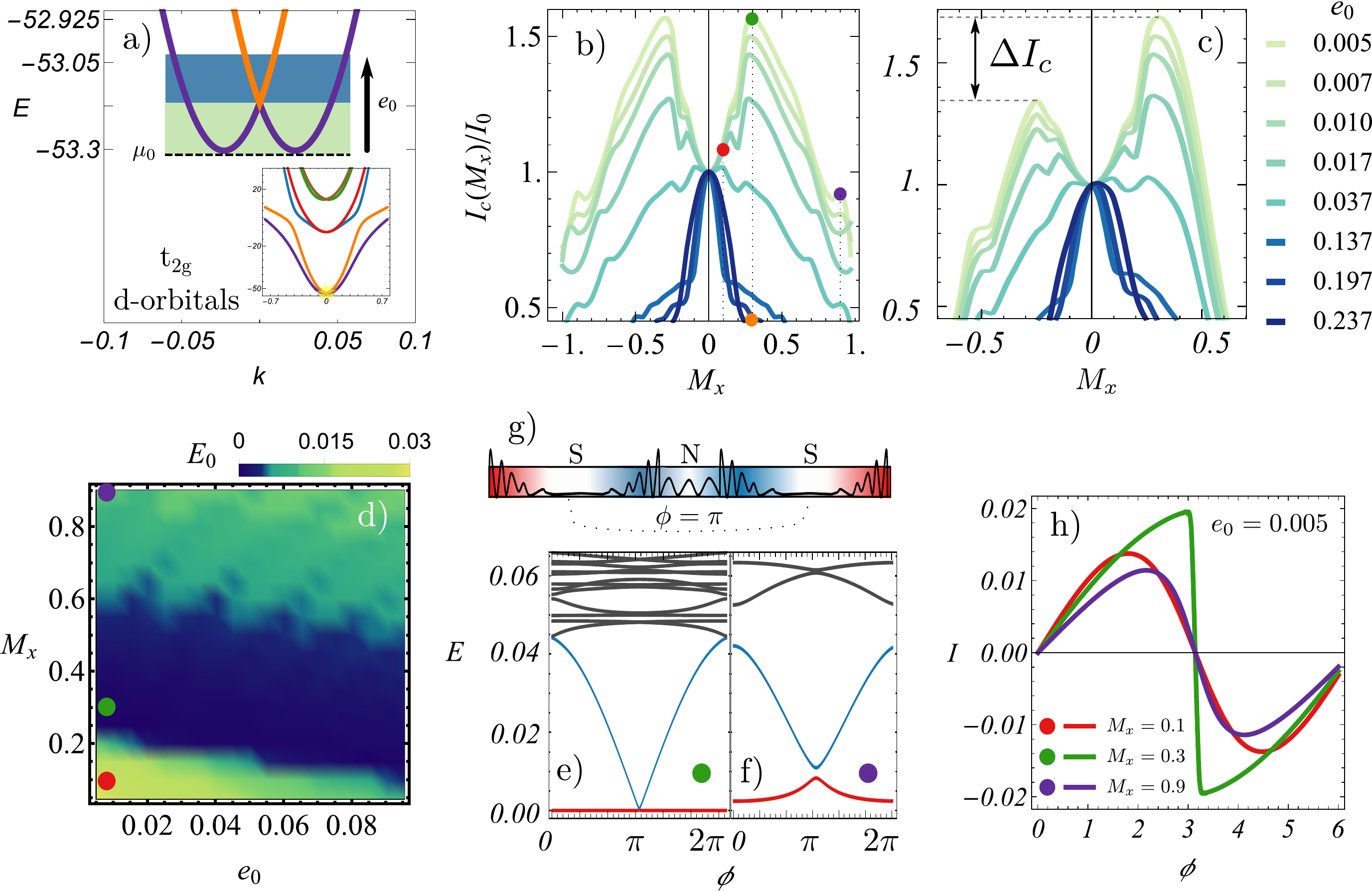}
	\end{center}

	\caption{a) Sketch of band structure nearby the $\Gamma$ point in the Brillouin zone for $M_x=0$. The six $t_{2g}$ d-orbitals, i.e. $d_{xy}$-like (purple and orange), $d_{zx}$-like (red and blue) and $d_{yz}$-like (green and brown), are reported in the inset. The two-tone shaded area (green and blue) shows the lowest doublet. b) Critical current in the SNS junction as a function of $M_x$ for different fillings factor $e_0$ and $q=0$. A strong enhancement of $I_c$ occurs at $|M_x| \simeq 0.3$ and low fillings ($e_0 < 0.137$). For $e_0 \geq  0.137$, the critical current rapidly decrease with increasing of $M_x$. c) Critical current in the SNS junction as in the previous panel and $q=\eta M_x$, with $\eta \sim 0.01$. An asymmetry at changing the field direction is observed. d) Phase diagram of the junction realized by plotting the lowest energy level of the half positive sector of the BdG spectrum for $\phi=\pi$, by varying $M_x$ and $e_0$. Zero-energy region (blue region) corresponds to the emergence of MBS within each superconductor, the range of $e_0$-values corresponds to the green shaded area of panel a) (low filling). e), f) Low-energy spectrum as a function of the superconducting phase difference $\phi$ for respectively $M_x=0.3$, $0.9$ as indicated by the corresponding pawns. It shows the presence of zero modes in the topological phase (panel e)), while the absence of MBSs in panel f) is clearly traced back to panel d). g) Sketch of the MBSs wavefunctions in the topological region for $\phi=\pi$, blue and red areas indicate the Majorana polarization. h) CPRs evaluated at the points indicated in panel d). CPRs alternate between sine-like behaviours (for $M_x=0.1$, $0.9$) and sawtooth profiles for $M_x=0.3$, in agreement with topological/trivial phases of the phase diagram (line's colors correspond to the points of panel b) and d)). CPRs fulfill the symmetry $I_{M_x}(\phi)=-I_{-M_x}(-\phi)$, see Appendix \ref{AppE}.}
	\label{Figure1}
\end{figure*}
$H=H^0+H^{SO}+H^{Z}+H^{M}$ with,
\begin{eqnarray}
&&	H^0\!\!=\!\!\sum_{j} \!\Psi^{\dagger}_j (h^{0}_{on} \otimes \sigma_0) \Psi_j\!+\!\Psi^{\dagger}_j (h^{0}_{hop} \otimes \sigma_0) \Psi_{j+1}\!+\!h.c.,\\
&&	H^{SO}\!=\! \Delta_{S0} \!\sum_{j} \Psi^{\dagger}_j \biggl(l_x \otimes \sigma_x + l_y \otimes \sigma_y + \l_z \otimes \sigma_z \biggr) \Psi_j,\\
&&	H^Z= -i \frac{\gamma}{2} \sum_{j} \Psi^{\dagger}_j \biggl(l_y \otimes \sigma_0 \biggr) \Psi_{j+1}+h.c.,\\
&&	H^M= M_x \sum_{j} \Psi^{\dagger}_j \biggl(l_x \otimes \sigma_0 + \l_0 \otimes \sigma_x \biggr) \Psi_j,
\end{eqnarray}
where we use the $t_{2g}$ orbitals ($d_{xy}$, $d_{yz}$, and $d_{zx}$), while $\Psi_j=(c_{yz,\uparrow,j},c_{yz,\downarrow,j},c_{zx,\uparrow,j},c_{zx,\downarrow,j},c_{xy,\uparrow,j},c_{xy,\downarrow,j})^T$ is a vector whose components are the electron annihilation operators for a given spin, orbital, and position. The Hamiltonian terms $H^0$, $H^{SO}$, $H^Z$, $H^M$ represent the kinetic energy, the spin-orbit, the inversion-symmetry breaking, and the Zeeman interaction term \cite{PhysRevB.100.094526}. We consider a nanochannel oriented along the $x$-direction, for which the topological phase is stabilized by a magnetic field perpendicular to the orientation of the orbital Rashba-like field \cite{PhysRevB.102.224508,PhysRevB.100.094526}.
$\sigma_i$ are the Pauli matrices, while $\sigma_0$ indicates the identity matrix. $l_x$, $l_y$, $l_z$ are the projections of the $L=2$ angular momentum operator onto the $t_{2_g}$ subspace. The analytic expressions of such matrices with those of the hopping Hamiltonians $h^0_{on}$ and $h^0_{hop}$ are reported in Appendix \ref{AppA}. According to the ab-initio estimates and on the basis of spectroscopic studies \cite{Pai_2018,PhysRevB.87.161102,Joshua12}, in agreement with Ref.\cite{PhysRevB.100.094526}, we assume: $t_1=300$, $t_2=20$, $\Delta_{SO}=10$, $\Delta_t=-50$, $\gamma=40$ (in unit of meV). $t_1$ and $t_2$ are the $x$-directed intraband hopping couplings, respectively for the $yz$ and $zx/xy$ bands, appearing in $H^0$. $\Delta_t$ denotes the crystal field potential induced by a symmetry lowering from cubic to tetragonal. The latter is also related to inequivalent in-plane and out-of-plane transition metal–oxygen bond lengths, which lowers the on-site energy of the $xy$-band. The lattice constant has been fixed at $a=3.9 \AA$. This set of parameters is representative of a physical regime with hierarchy of the electronic energy scales such that $|\Delta_t|>\gamma>\Delta_{SO}$.\\
In order to introduce superconducting correlations and define the SNS junction geometry, we add to the normal part of the Hamiltonian $H$ the mean-field pairing contribution:
\begin{eqnarray}
	H^P = \sum_{j,\alpha} \Delta(j,q)\ c^{\dagger}_{j,\alpha,\uparrow} c^{\dagger}_{j,\alpha,\downarrow} + h.c.,
\end{eqnarray}
where $\alpha$ stands for an orbital index and $\Delta (j,q)=\Delta e^{iqj} e^{i\phi_j}$ is a space-dependent gap, taking into account a phase gradient ($\phi_j=\chi_j^L \phi_L+\chi_j^R \phi_R$, with $\chi_j^\alpha=1$ only when $j$ belongs to the $\alpha=L, R$ electrode) induced by the bias current and a finite momentum effect of the Cooper pair ($q$). Indeed, for some classes of superconductors, where both inversion and time-reversal symmetries are broken, and in the presence of antisymmetric SOC, it has been shown that Cooper's pairs acquire a finite momentum  \cite{doi:10.1073/pnas.2119548119}. Therefore, we have included a spatially modulated superconducting gap, with $q$ continuously determined by the magnetic field ($q=\eta M_x$). The pairing amplitude for both superconductors has been fixed as $\Delta=0.05$ \cite{PhysRevB.102.224508}, while the phase gradient is given by $\phi=\phi_L-\phi_R$, being $\phi_{L,R}$ the phase of the order parameter of the left and right lead.\\
The Josephson current of a short SNS junction with translational invariant leads can be efficiently calculated by using the subgap Bogoliubov-de Gennes (BdG) spectrum of a system with truncated superconducting leads as $I(\phi)=-(e/\hbar) \sum_{n} dE_n/d\phi$  \cite{PhysRevB.96.205425}, with $E_n$ being the subgap energies of the BdG spectrum. The finite lead approach is valid as long as the short junction limit is considered ($L_N \ll \xi$, with $\xi$ the Bardeen-Cooper-Schrieffer (BCS) coherence length). Thus, we set the total size of the system, $L=2L_S+L_N$, with $L_S$ being the size of the superconducting lead and $L_N \ll L_S$ the normal region size. The maximum (absolute value of minimum) of $I(\phi)$ yields the critical current in the positive (negative) direction.
The tight-binding Hamiltonian is numerically treated by using KWANT \cite{Groth_2014} and solved with the help of NUMPY routines \cite{citeulike:9919912}. We explore different electronic regimes defined by the orbital filling and controlled by $\mu$, also varying the Zeeman energy $M_x$, and the phase difference $\phi$. Numerical simulations have been performed by presenting the chemical potential in the form  $\mu=\mu_0+e_0$, where $e_0$, which determines the orbital filling, represents the energy offset measured from the bottom $\mu_0$ of the considered band. Appropriate setting of $\mu_0$ allows to characterize the orbital-sensitive response of the system.

\begin{figure}
	\centering
	\includegraphics[width=0.48\textwidth]{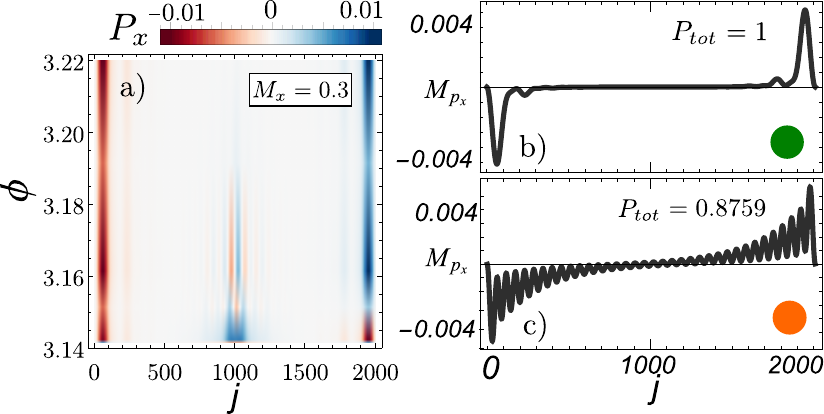}

	\caption{a) $P_x$ at $\phi=\pi$ shows the presence of four MBS dislocated at the corners of the two superconductors. For $\phi \gtrsim \pi$ the inner Majoranas disappear, and the whole topological charge concentrates at the corners, see Appendix \ref{AppD}. b), c) Majorana polarization $P_x$ for $\phi=\pi/40$ and for $M_x=0.3$ of the lightest green and darkest blue curve of Fig.~\ref{Figure1}b), as indicated by the corresponding points. The insets, $P_{tot}$ quantifies the topological charge of the exhibited modes.}
	\label{Figure2}
\end{figure}

\section{Numerical results} 
First, we focus on the regime of orbital filling corresponding the lowest doublet of the energy spectrum with $d_{xy}$ orbital character (see Fig.~\ref{Figure1}a). In Fig.~\ref{Figure1}b),c) we report the critical current pattern by varying the Zeeman energy for both $q=0$ and $q=\eta M_x$, with $\eta \simeq 0.01\ meV^{-1}$, deduced by data in Ref. \cite{NPJ066}. Hereafter, we consider an extended Zeeman energy range that is accessible without destroying the superconductivity \cite{Hc,Hc2}, because the effective Landé g-factor can be strongly amplified by confinement effects \cite{PhysRevB.54.R14257,PhysRevMaterials.4.122001}. We notice that, when $e_0$ is varied in the green shaded area in Fig.~\Ref{Figure1}a), an anomalous supercurrent enhancement occurs with increasing the Zeeman energy $M_x$. When $e_0$ fails in the blue region in Fig.~\ref{Figure1}a), a lowering of the critical current for increasing $M_x$ is observed, like in the conventional spin-singlet superconductivity case. The anomalous enhancement of the critical current is a consequence of the spin-momentum locking phenomenon which is realized when the chemical potential lies at the bottom of the Rashba-like band (purple lines in Fig.~\ref{Figure1}a)). On the other hand, at increasing filling, imperfect spin-momentum locking is realized, so that the supercurrent decreases with the magnetic field. Further increasing $e_0$, the spin-momentum locking can be completely lost when the Fermi energy level cuts the two spin-subbands, (purple and orange lines in Fig.~\ref{Figure1}a)). In this case, spin and momentum degrees of freedom are decoupled.\\ 
\begin{figure}
    \centering
	\includegraphics[width=0.4\textwidth]{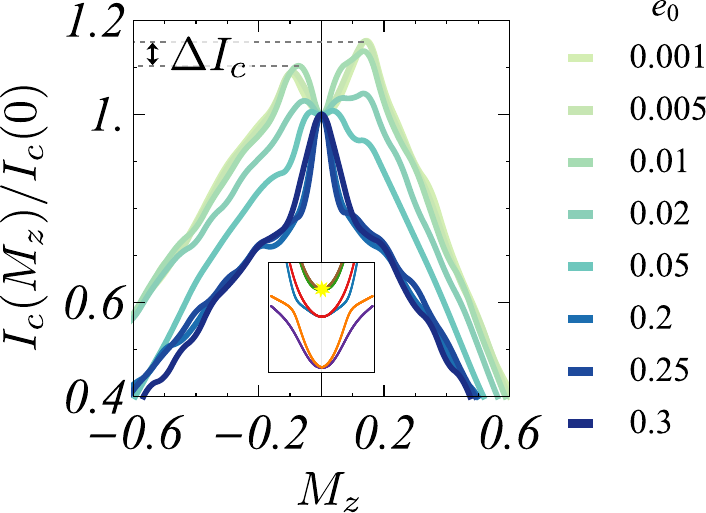}\\
	
	\caption{a) Critical current in the SNS junction as a function of $M_z$ for different fillings factor $e_0$ starting from the bottom of the highest doublet in the spectrum (see inset) and $q=\eta M_z$, with $\eta=0.005$ . The asymmetric peaks collapse in a single maximum at higher fillings.}
	\label{Figure3}
\end{figure}
\begin{figure*}
	\begin{center}
		\includegraphics[scale=0.22]{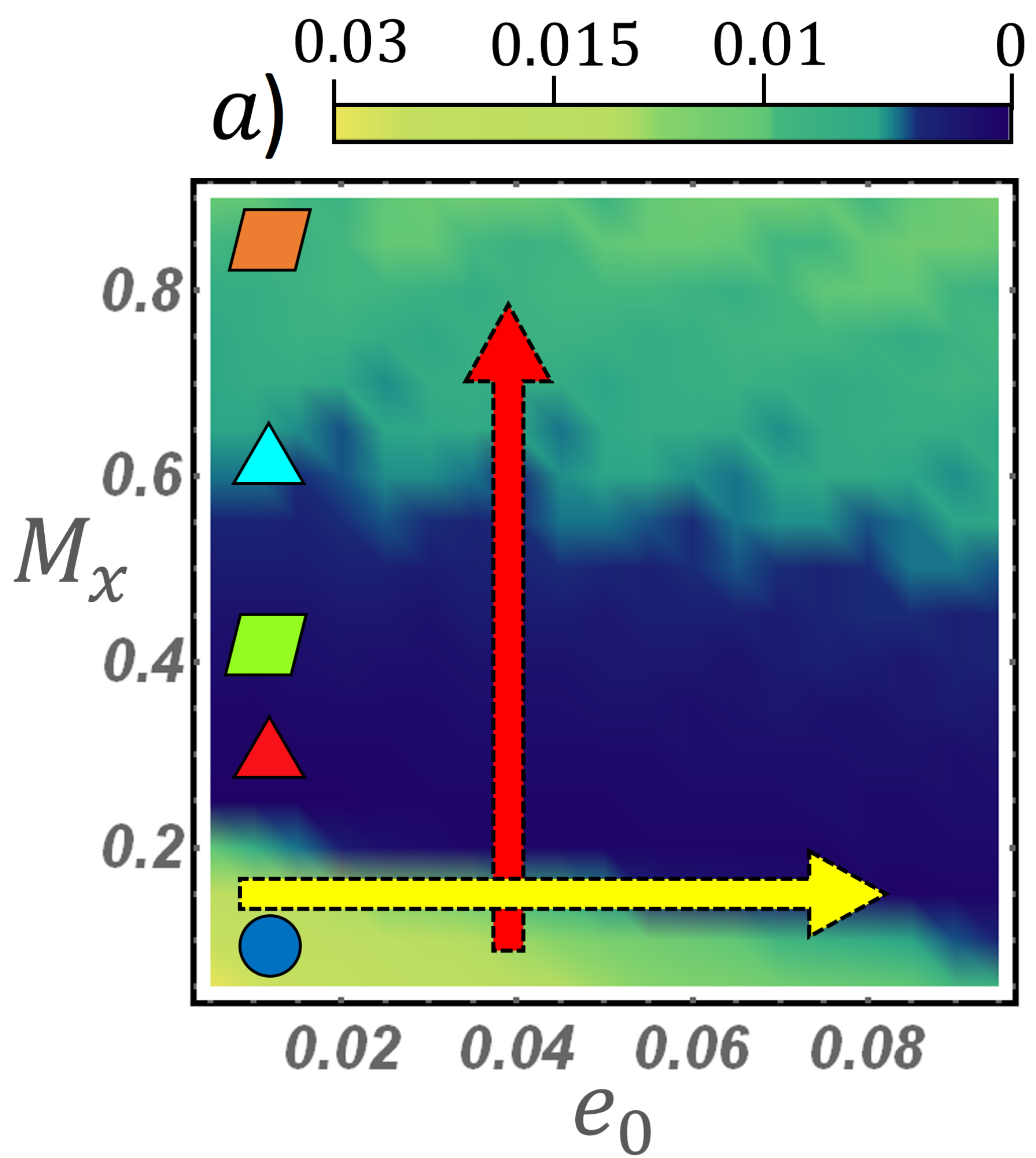}
		\hspace{0.5cm}
		\includegraphics[scale=0.22]{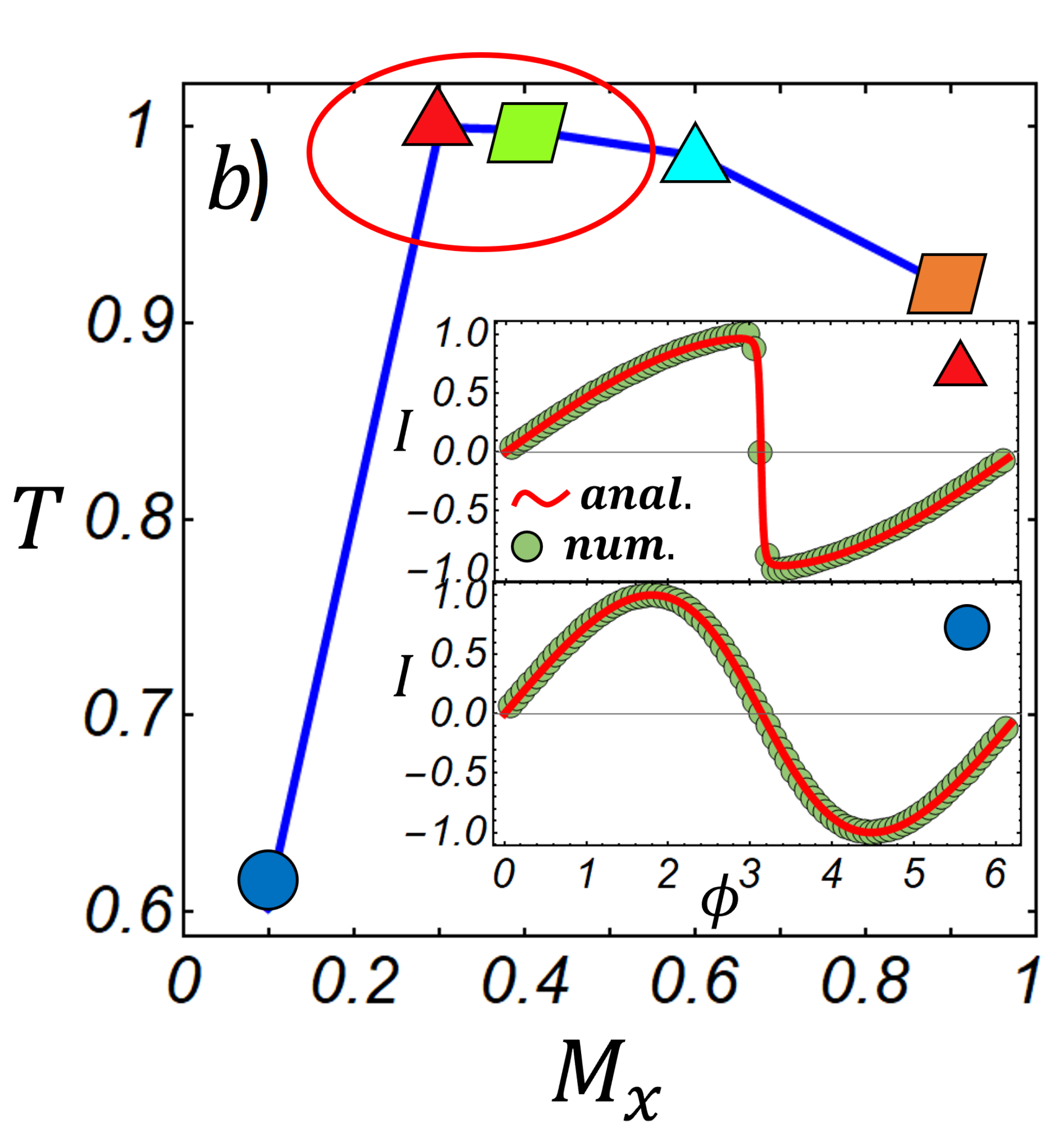}\\
	\end{center}
	\includegraphics[scale=0.2]{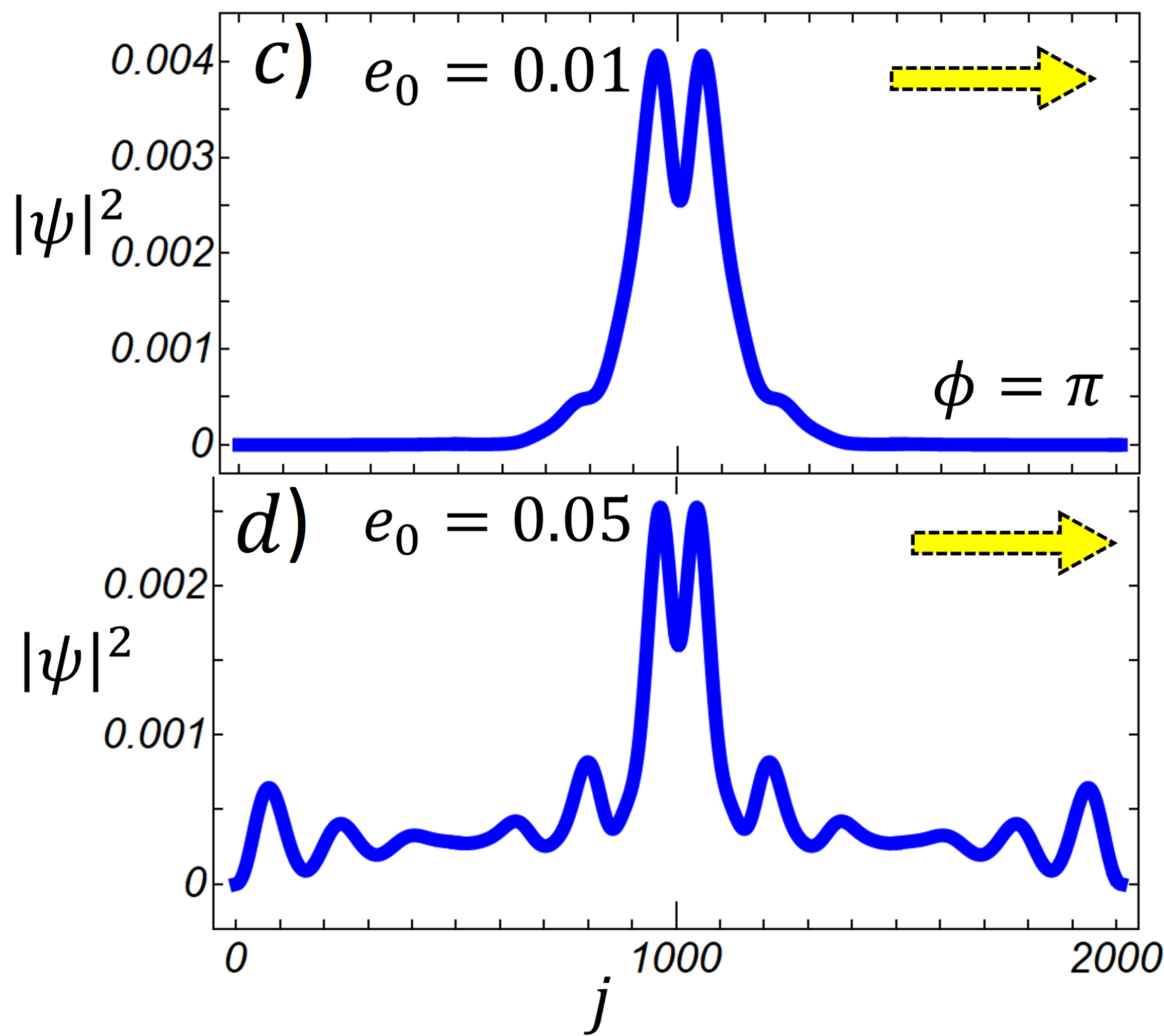}
	\includegraphics[scale=0.2]{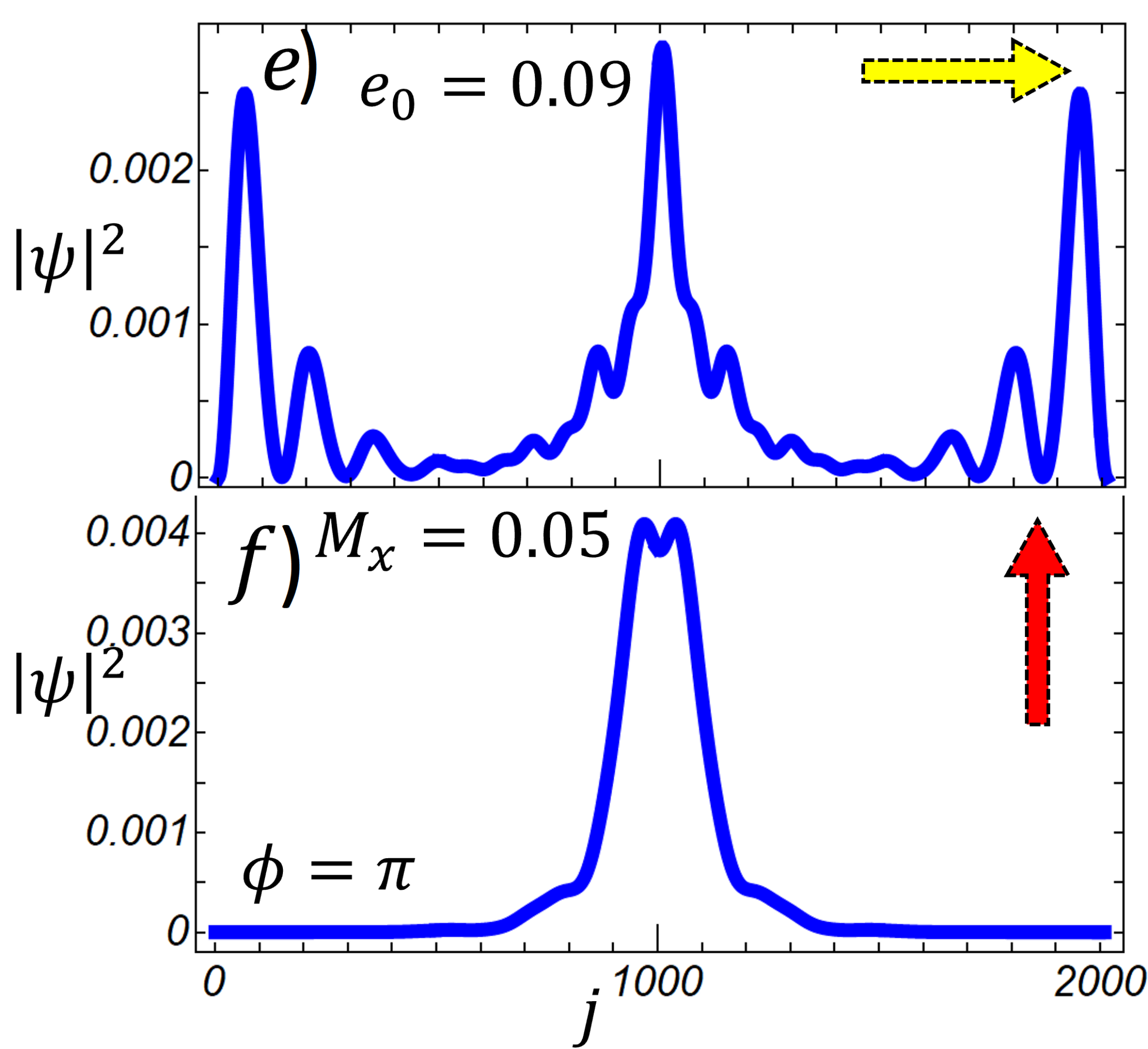}
	\includegraphics[scale=0.2]{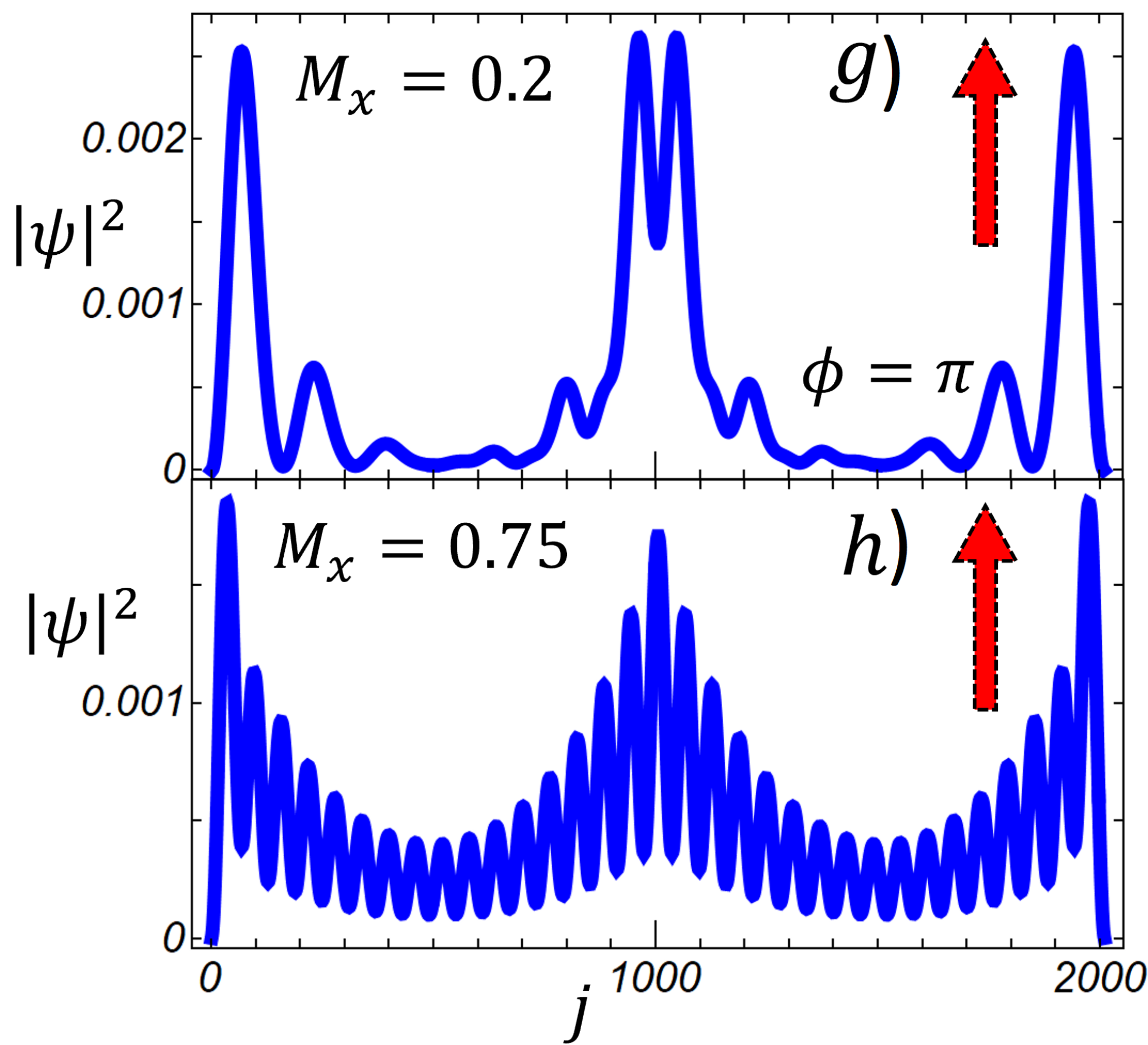}
	\caption{a), The phase diagram of the main text has been reported with the two coloured arrows indicating directional cuts. b), By following the vertical pawns of panel a), we plot the effective transparency of the $SNS$ junction extracted by fitting the analytical expression of $I(\phi)$ in presence of bound states, Eq. \ref{T}, and the data coming from numerical diagonalization of the Hamiltonian. Examples of the agreement between analytic curves and numerical data are reported in the inset of the panel. MBSs are associated to a fully transparent junction $T=1$ (pawns in the blue region of phase diagram), while for hybridized modes $T$ progressively decreases (greenish yellow area of phase diagram). The minimum of $T$ is associated to the pawn in the yellow region of panel a where no Majorana fermions are hosted by the system. Panels from c) to h) show the square modulus of the lowest energy eigenstates following the directional coloured arrows of a). Trivial modes, Majorana fermions and hybridized states can be easy recognized when the system belongs respectively to yellow , blue and greenish yellow regions of the phase diagram. For $\phi=\pi$ a single or double headed central peak can be observed, related to the inner Majorana states of the superconducting leads.}
	\label{SuppMat1}
\end{figure*}
Spin-momentum locking is one of the most important property associated with non-trivial topological states \cite{TIAN20141,D0NR06590K}, and a topological phase transition is expected when $e_0$ lies at the bottom of the Rashba-like band and $M_x$ increases above a critical magnetic field. The latter is signaled by a gap closing and reopening with the formation of MBSs at zero energy.  The density plot in Fig.~\ref{Figure1}d) captures this phenomenology by showing the behaviour of the lowest energy eigenvalue $E_0$ of the Hamiltonian as a function of $e_0$ and $M_x$. The Andreev spectrum as a function of the superconducting phase is shown in Fig.~\ref{Figure1}e),f) for two points of panel b). We see a gap closing in e) indicating a topological phase transition. Indeed, in this regime, the lowest level (red line) is almost insensitive to $\phi$ and comes from the outer MBSs of the two superconducting leads. The second energy level (blue line) originates from the inner Majoranas (shown in the panel g)). It is strongly dispersive with $\phi$ and becomes degenerate with the lowest energy level only at $\phi=\pi$. Outside the topological region both the first and the second energy levels are lifted from zero (Fig.~\ref{Figure1}f)). For $\phi\neq\pi$, the inner MBSs hybridize into a fermionic state of finite energy, as discussed below. Fig.~\ref{Figure1}h) displays the CPRs in the cases highligthed in Fig.~\ref{Figure1}d). They show sawtooth profiles when $E_0=0$, while, in the opposite case, $E_0 \neq 0$, they acquire a sine-like behaviour as a function of $\phi$ \cite{PhysRevB.93.220507}. Introducing an effective junction transparency $T$ depending on the value of $E_0$, we observe that skewed CPRs correspond to $T \rightarrow 1$ and resonant transmission with MBSs, while $E_0>0$ gives sine-like CPRs and $T < 1$ (details at Appendix \ref{AppB}). \\
The superconducting topological phase can be well characterized by the 
Majorana polarization ($M_p$), which measures the quasiparticle weight in the Nambu space. According to Refs. \cite{BENA2017349,PhysRevB.102.085411,PhysRevB.102.224508,MaiellaroEPJplus}: $P(j,\omega)=\sum_{n}\bigl( \sum_{\alpha,\sigma} u_{n,j,\alpha,\sigma} v_{n,j,\alpha,\sigma}\bigr) (\delta(\omega-E_n)+\delta(\omega+E_n))$, with $u$ and $v$ the particle and hole components of the BdG wavefunction,  while
the indices $\alpha$ and $\sigma$ are related to the orbital and spin degrees of freedom. In particular, by choosing $\omega=0$, the total Majorana polarization $P_{tot}=|\sum_{j=1}^{L/4} P(j,0)|$ is equal to $1$ for genuine MBSs. $P$ defines a vector with $P_x = Re[P]$ and $P_y = Im[P]$, and both $P_x$ and $P_y$ are peaked functions at the system edges, i.e. they show MBSs with opposite topological charge. In particular, Fig.~\ref{Figure2}a) shows the appearance of four MBSs for $\phi=\pi$, while for $\phi \sim \pi+\varepsilon$ (i.e. we set $\varepsilon=0.05$),  the inner MBSs are fully hybridized and Majorana polarization disappears at the middle of the junction, see Appendix \ref{AppD}. When $\phi= \pi/40$, fully-polarized MBSs nucleate at the system edges (panel b)), while hybridized modes show a polarization loss (panel c)).
The scenario described above suggests that the symmetric enhancement of the critical current vs magnetic field curves, reported in Ref. \cite{NatCommun14984}, is a fingerprint of a topological phase transition.\\ 
The same type of analysis based on the Majorana polarization holds also in the case of  Fig.~\ref{Figure1}c), where an asymmetric enhancement of $I_c$ vs magnetic field is observed. This behavior closely follows the experimental findings of the oxide-based Josephson junction \cite{deluca_2014, NPJ066}\footnote{Both Refs. \cite{NatCommun14984,NPJ066} consider a magnetic field along the $z$-direction. We have verified that changing the magnetic field direction from $x$ to $z$  has no physical significance in the emerging phenomenon and the latter considerations still remains true. However, in view of the non-trivial multiband effects, small quantitative differences are shown for $M_i \geq 0.3$ ($i=x,z$). In Appendix \ref{AppF} we compare the two directions of magnetic field.}.\\  
Phenomenologically, the asymmetry was ascribed to the inversion symmetry breaking and to the presence of three current channels in the junction, with phases 0, $\pi$ and $\phi_0$ \cite{NPJ066}.The analysis of the CPRs for the two asymmetric maxima still shows a sawtooth profile when the magnetic field approaches the topological phase transition. Interestingly, the critical current pattern is antisymmetric with respect to the inversion of both the magnetic field and the polarisation current, see Appendix \ref{AppE} for more details. Thus, the formation of a $\phi$-junction and its relevant role for the system response suggested in Ref. \cite{NPJ066} is here confirmed.\\
Finally, in order to explore an alternative filling regime, we have studied the critical current patterns (see Fig.~\ref{Figure3}) for chemical potentials corresponding to the upper bands doublet in Fig.~\ref{Figure1}a)). These bands originate from the hybridization of orbitals with $d_{zx}$, $d_{yz}$ character, with a Rashba-like dispersion. As previously shown \cite{PhysRevB.102.224508}, this band exhibits a transition to the topological superconductivity when a  magnetic field is applied along the $z$-axis. The critical current pattern shows again a double maxima structure that coalesce in a single maximum when the energy offset $e_0$ is moved from the bottom of the doublet to higher values. The asymmetric increase of the current pattern with the applied magnetic field still signals the approach to a topological phase transition with the formation of MBSs at the edges of the 1D channel. 
\section{Conclusions.} 
In summary, we have highlighted that signatures of topological superconductivity can be found in the anomalous critical current pattern of short superconductor-normal-superconductor junctions based on oxide nanochannels and, in general, on non-centrosymmetric superconductors. 
Notably, the topological phase transition is signaled by an enhancement of the critical current by increasing the applied magnetic field perpendicular to the spin-orbit coupling. These properties are compatible with the presence of MBSs at the edges of the superconducting leads and the topological properties are robust to multiband effects whose relevance can be tuned by appropriate gating of the system.\\
Finally, the microscopic phase transition mechanism, reported in this letter, appears to be consistent with recent experimental observations of unconventional features of the Josephson current, for which, to date, a microscopic theory is still lacking. 
\section*{acknowledgments}
\vspace{-0.6cm}
We acknowledge A. Kalaboukhov, G. Singh, M. Salluzzo and M. Cuoco for fruitful discussions.
\appendix
\section{Kinetic and orbital momentum matrices of the oxide nanochannel}
\label{AppA}
\vspace{-0.5cm}
We report the matrix expressions for Kinetic Hamiltonians $h^{0}_{on}$, $h^{0}_{hop}$ and for the angular momentum projections $l_x$, $l_y$ and $l_z$ for the oxide nanochannel:
\begin{eqnarray}
	&&h^{0}_{on} = \left( \begin{array}{ccc}
		2 t_2 -\mu & 0 & 0\\
		0 & 2 t_1-\mu & 0 \\
		0 & 0 & 4 t_1+\Delta_t-\mu \end{array} \right),\\[10pt]
	&&h^{0}_{hop} = - \left( \begin{array}{ccc}
		t_2  & 0 & 0\\
		0 &  t_1 & 0 \\
		0 & 0 &  t_1 \end{array} \right),\ l_x = \left( \begin{array}{ccc}
		0  & 0 & 0\\
		0 &  0 & i \\
		0 & -i &  0 \end{array} \right)\\[10pt]
	&& l_y= \left( \begin{array}{ccc}
		0  & 0 & -i\\
		0 &  0 & 0 \\
		i & 0 &  0 \end{array} \right),\ l_z = \left( \begin{array}{ccc}
		0  & i & 0\\
		-i &  0 & 0 \\
		0 & 0 &  0 \end{array} \right).
\end{eqnarray}
\section{topological phase and effective transparency}
\label{AppB}
The SNS junction experiences a topological phase transition for low fillings $e_0$ of the $d_{xy}$ and $d_{yz}$ orbitals. In Fig.~\ref{SuppMat1} some additional results, corroborating the ones of the main text, are reported. In particular, by following the vertical pawns of the phase diagram we can compute the emerging effective transparency of the junction. Indeed, for a Josephson junction much shorter then the superconducting coherence length, a well-known relation exists between the supercurrent $I$, the phase difference $\phi$ and the transparency of the junction $T$ \cite{PhysRevLett.124.226801}:
\begin{eqnarray}
	I= \frac{e \Delta}{\hbar} T \frac{\sin(\phi/2) \cos(\phi/2)}{\sqrt{1-T \sin^2(\phi/2)}}.
	\label{T}
\end{eqnarray}
In Fig.~\ref{SuppMat1}b) we show that an effective transparency $T=1$ corresponds to the non-trivial points of the phase diagram. While for points in the yellow and greenish yellow regions $T$ is greatly reduced. By following the directional cuts in panel a) we also report the evolution of the square modulus of the lowest energy modes of the BdG spectrum. In particular, for both the horizontal and vertical cuts, trivial modes, Majorana fermions, and hybridized states can be easily recognized when the system belongs respectively to yellow, blue, and greenish yellow regions of the phase diagram. For $\phi=\pi$ a single or double headed central peak can be observed, indicating the presence of inner Majorana states of the superconducting leads.
\section{Some additional results on critical current}
\label{AppC}
\begin{figure*}
	\includegraphics[scale=0.24]{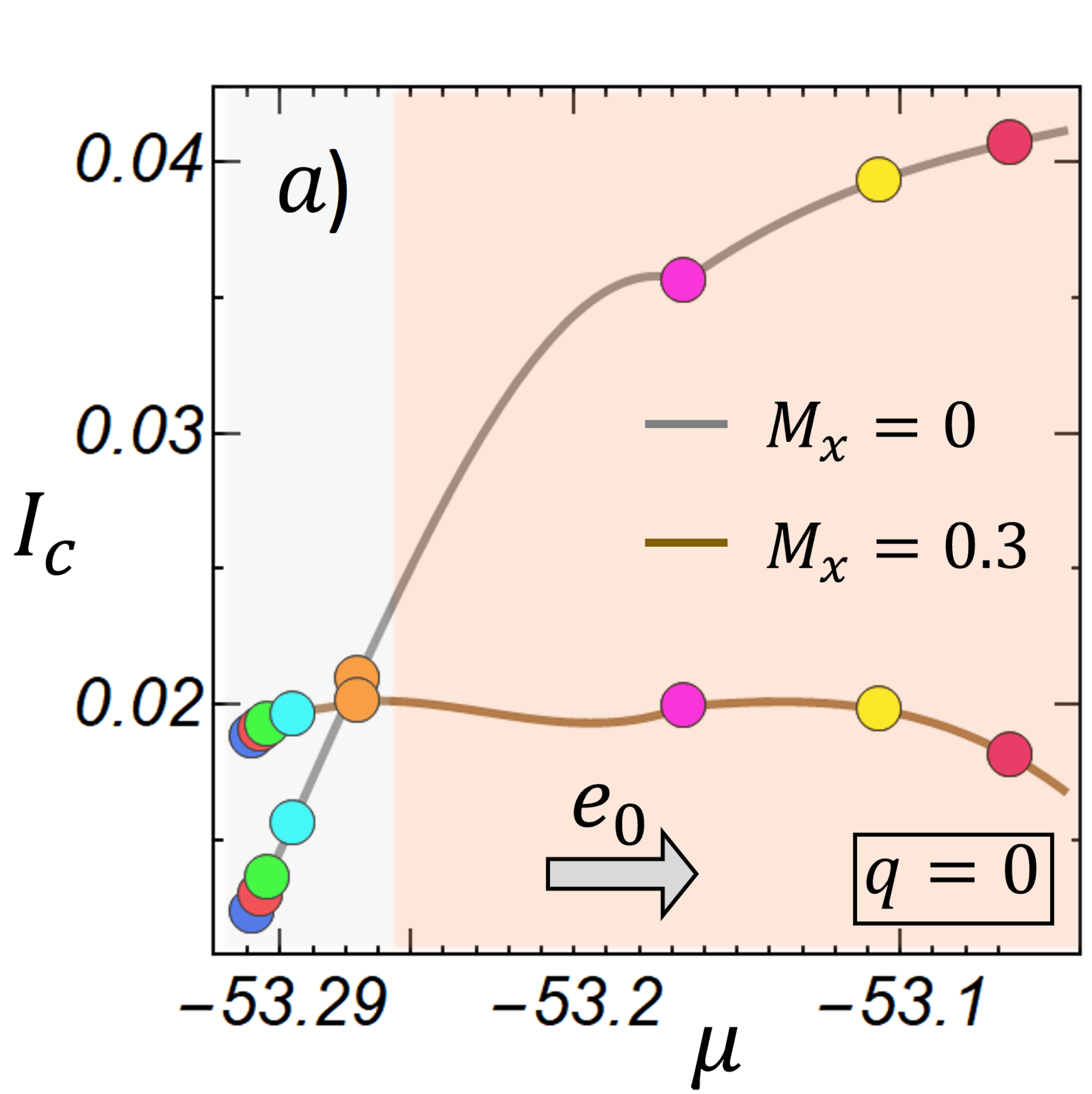}
	\hspace{0.2cm}
	\includegraphics[scale=0.24]{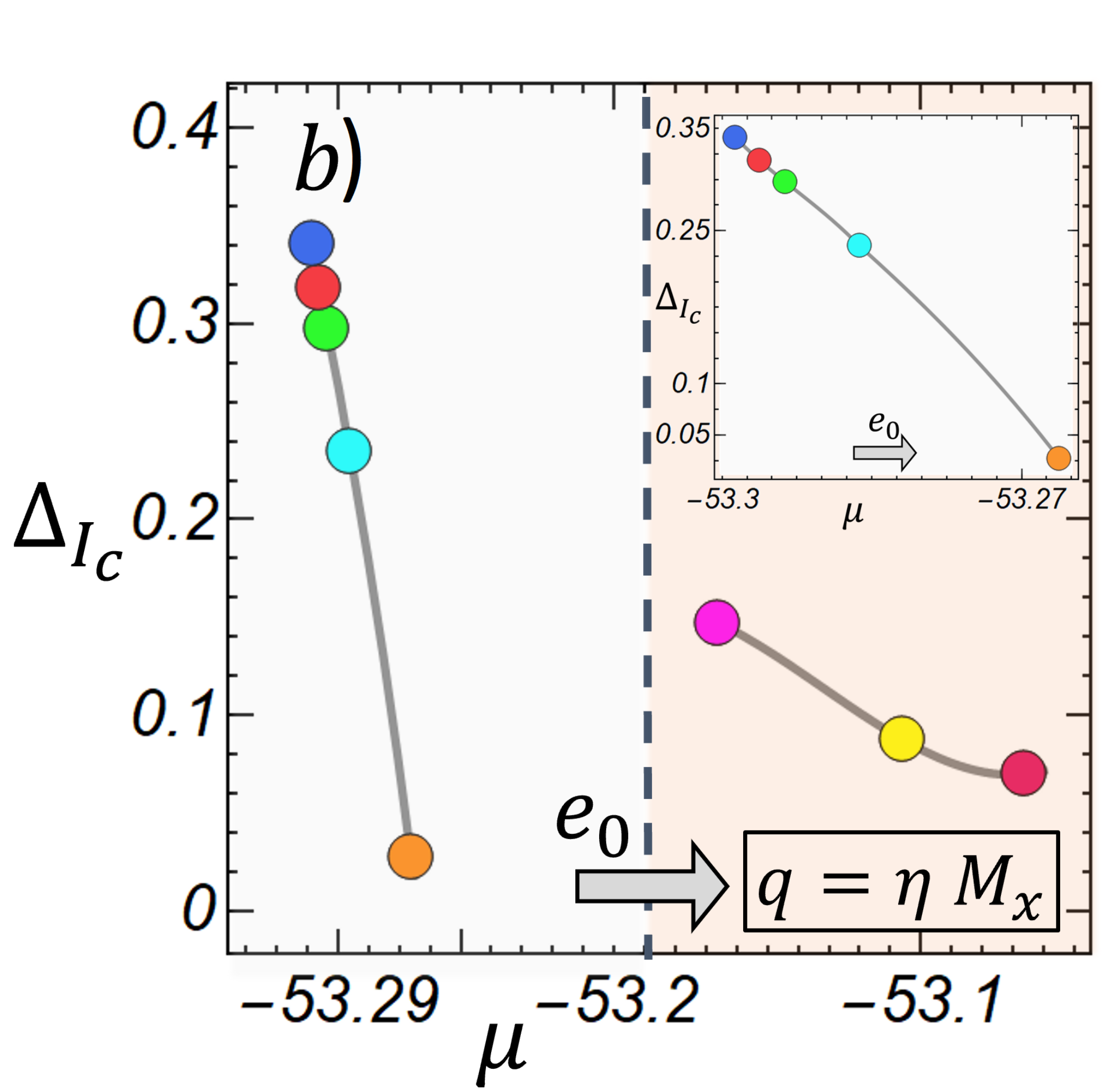}
	\hspace{0.2cm}
	\includegraphics[scale=0.22]{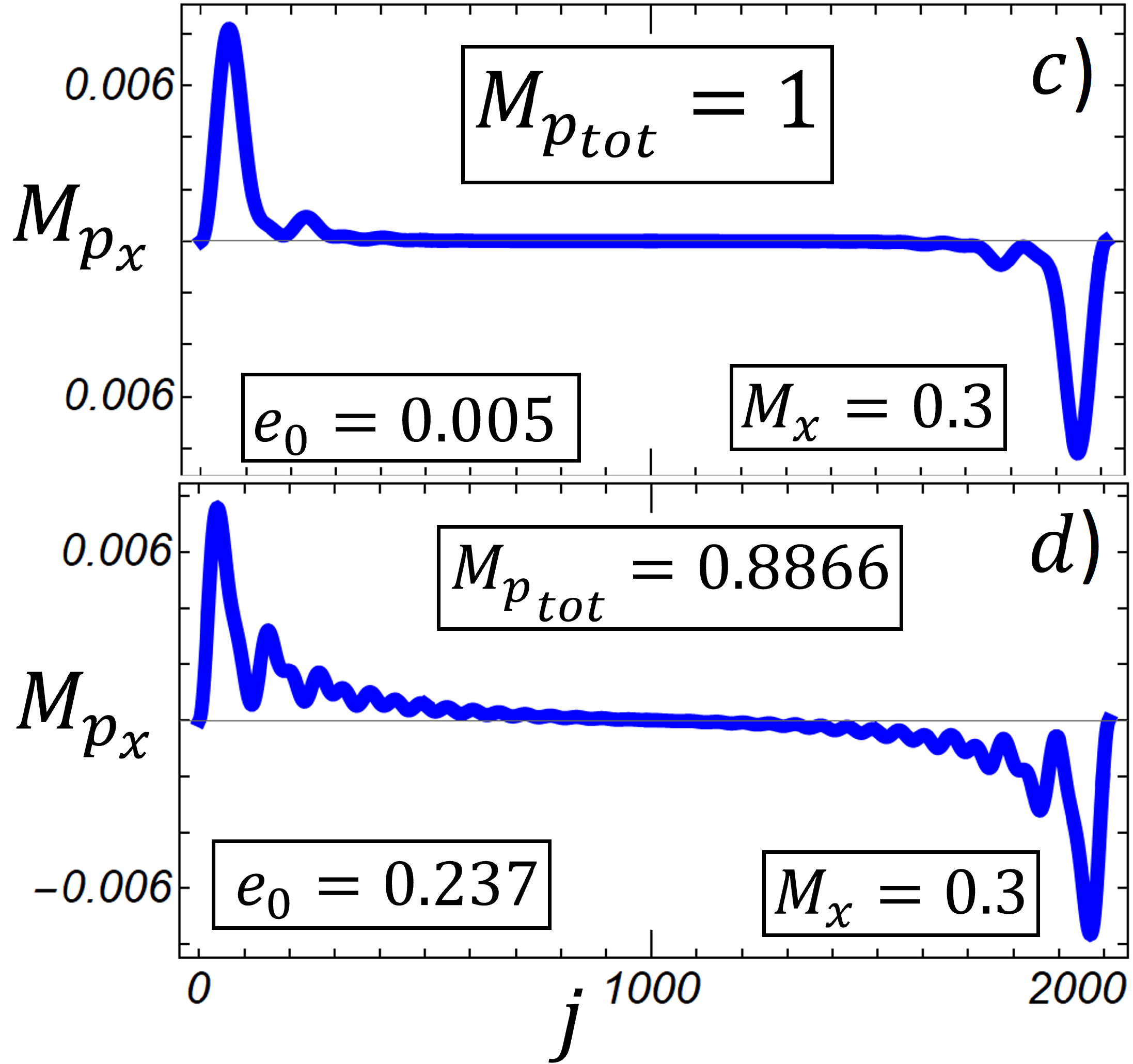}
	\caption{a), $I_c$ for increasing filling factors $\mu$ by fixing $M_x=0$, $0.3$ when $q=0$. b), The asymmetry decreases monotonously for increasing $e_0$. c) and d) show $M_{p_x}$ for curves in Fig.~\ref{Figure1}c) of the main text obtained by setting $e_0=0.005$ and $e_0=0.237$, respectively, and $M_x=0.3$. In such cases the x-component of local Majorana polarization shows the two outer MBS with $M_{p_{tot}}=1$ in panel c) and a degree of hybridization between MBS with $M_{p_{tot}}\sim 0.88$ in panel d).}
	\label{SuppMat2}
\end{figure*}
In Fig.~\ref{SuppMat2}a) we compare $I_c$ as a function of the fillings $\mu$ estimated at zero field ($M_x=0$) and at the maximum ($M_x=0.3$) of the curves in Fig.~\ref{Figure1}b) of the main text. It is readily evident that the $I_c(0)$ shows more rapid suppression with $\mu$ in comparison to the smooth reduction at $M_x=0.3$. When the spatial gap modulation is present, see Fig.~\ref{SuppMat2}b), we observe that for the same fillings of panel a) the asymmetry monotonously decreases as the fillings are increased. Finally, panels c) and d) show that beyond the asymmetry, genuine MBSs are still related to the maximum of the $I_c$ curves with sawtooth CPRs. Indeed, in Fig.~\ref{SuppMat2}c) $M_{p_x}$ is well localized at the superconductors edges with $M_{p_{tot}}=1$ in correspondence of the maximum of critical current curve of the main text with $e_0=0.005$. A degree of hybridization appears in panel d) with $M_{p_{tot}}\sim 0.88$, referring to Fig.~\ref{Figure1}c) of the main text with $e_0=0.237$.
\begin{figure}
	\includegraphics[scale=0.2]{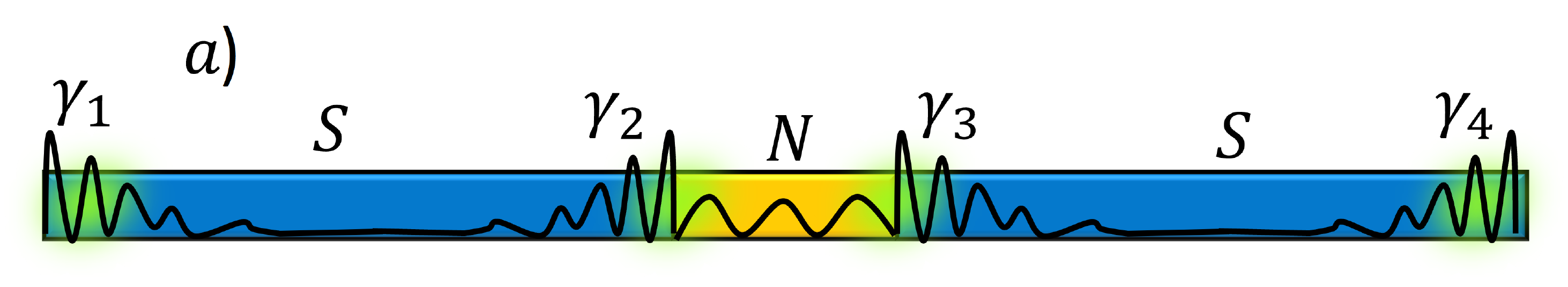}\\
	\includegraphics[scale=0.25]{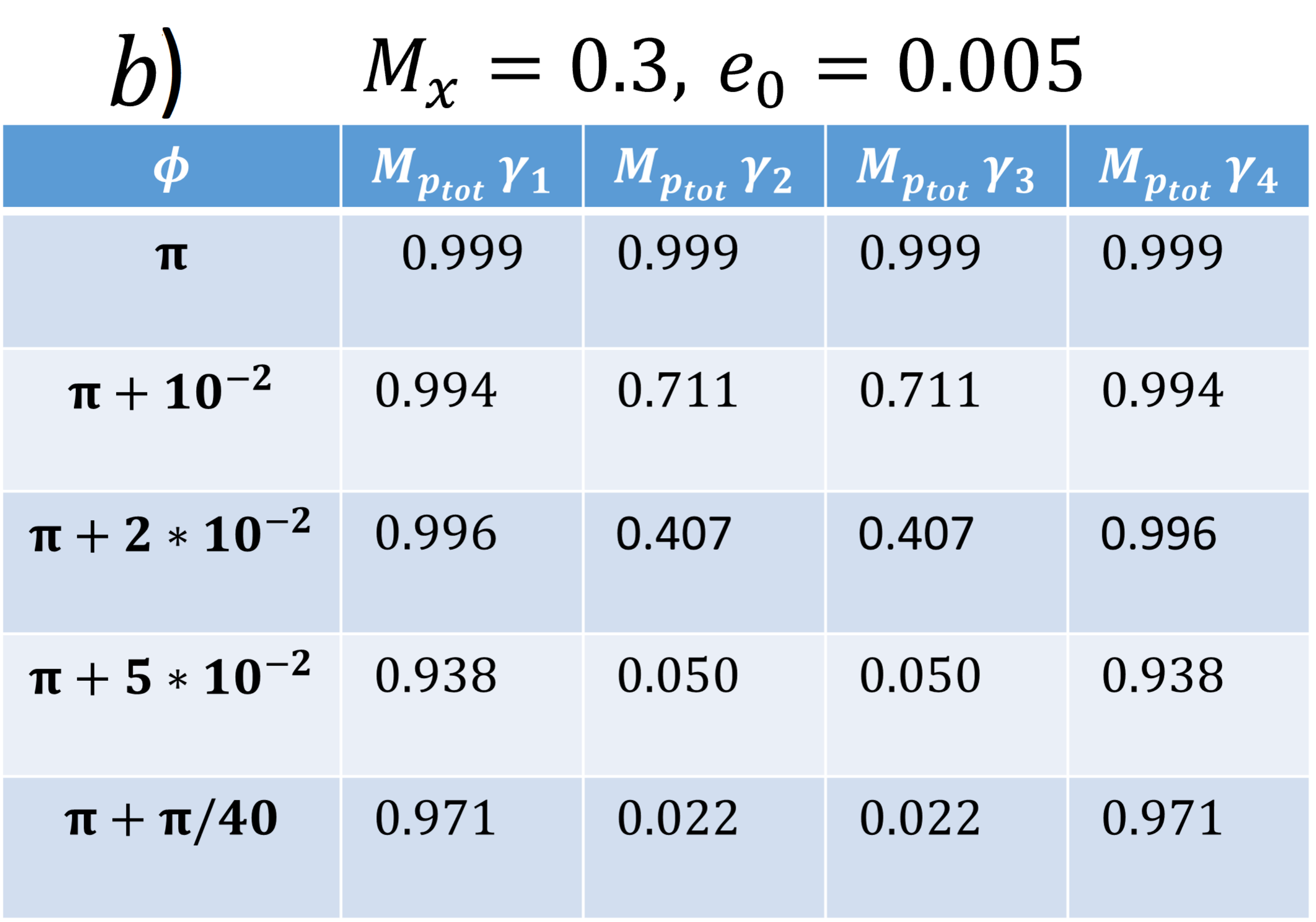}
	\caption{a), sketch of the junction hosting four MBSs for $\phi \simeq \pi$. b, Total Majorana polarization of the four symmetric lowest-energy states in a topological point $M_x=0.3$, $e_0=0.005$ and nearby $\phi=\pi$. For $\phi=\pi$ the system admits two inner and two outer Majorana states within the superconducting regions. As we shift from such point the inner Majoranas progressively hybridize, being significantly deviated form Majorana charge condition for $\phi \simeq\pi+5*10^{-2}$. Outside such range only the outer Majoranas survive.}
	\label{SuppMat3}
\end{figure}
\begin{figure*}
	\includegraphics[width=\textwidth]{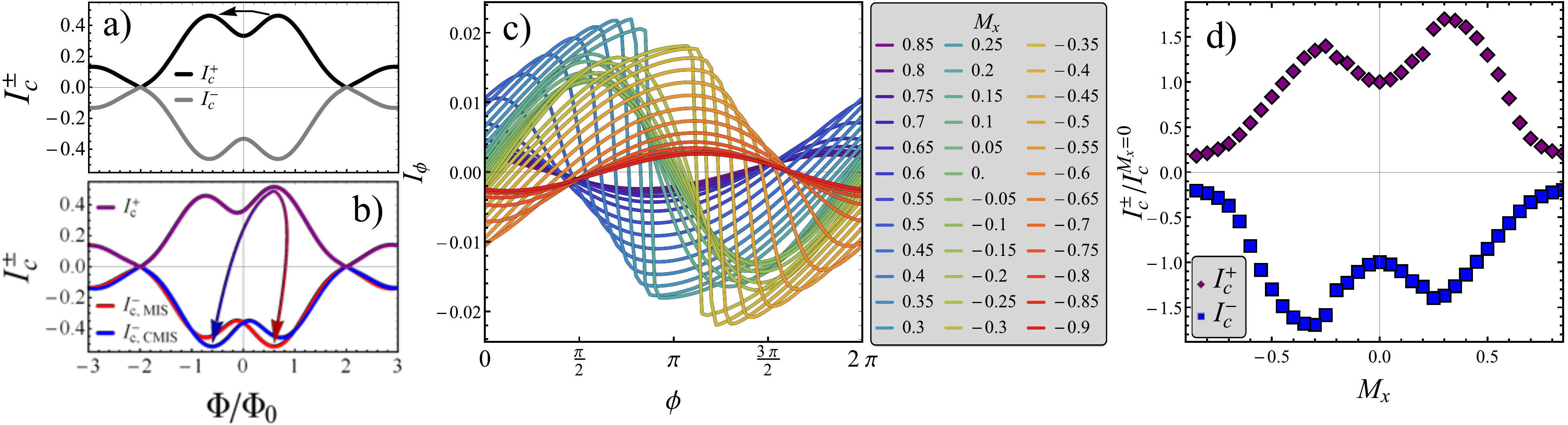}
	\caption{a)-b), Cartoon showing the $I_c^\pm(\Phi)$ patterns in the case of a) CIS, b) MIS (purple and red curves) and CMIS (purple and blue curves). The arrows indicate how a specific point (i.e., the top-right peak) of the pattern would reflect in the case of a reversal of the current and magnetic field. c) Collection of CPRs at different values of $M_x$. d) Switching currents as a function of $M_x$, for positive and negative direction of the bias current, assuming the CPRs showed in c).}
	\label{SuppMat5}
\end{figure*}
\section{Majorana polarization nearby $\phi=\pi$}
\label{AppD}
In table of Fig.~\ref{SuppMat3}, panel b), we quantify $M_{p_{tot}}$ for the four Majorana states hosted by the two superconducting leads nearby $\phi=\pi$ for a topological point of the phase diagram, $M_x=0.3$, $e_0=0.005$. For convenience the outer MBSs are indicated by $\gamma_1$, $\gamma_4$ and the inner by $\gamma_2$, $\gamma_3$. A sketch of the junction hosting the four MBSs is reported in panel a). We observe that the inner MBSs, $\gamma_2$, $\gamma_3$, progressively hybridize when $\phi$ deviate from $\pi$, being significantly hybridized for $\phi=\pi+5\cdot10^{-2}$, $M_{p_{tot}} \sim 0.05$. The outer MBSs, $\gamma_1$, $\gamma_4$, remain robust for all the topological phase: $M_{p_{tot}}\sim 1$.
\section{Symmetry of the critical current pattern}
\label{AppE}
Here, we recall the general argument presented in Ref.~\cite{Kashiwaya19} in the case of a Josephson junction formed by unconventional superconductors, in order to reveal symmetry breaking through the Josephson effect. First, we need to bear in mind that the CPR, i.e. the Josephson current $I_{\phi}$, can be generally decomposed into harmonic terms~\cite{Golubov04} $I_\phi=\sum_{n}\left [ I_{c,n}^s\sin\left ( {n\phi} \right )+I_{c,n}^c\cos\left ( {n\phi} \right ) \right ]$, with $n$ being a positive integer. Time-reversal symmetry (TRS) breaking requires the cosine terms to be finite; in other words, finding TRS through the Josephson effect is akin of recognising the presence of cosine terms in the Josephson current. In this case, a non-zero Josephson current is expected at zero phase, i.e., $I_{\phi}\neq0$ at $\phi=0$.
In order to identify the symmetries of the system, the magnetic field response of the Josephson critical current (namely, the maximum value of $I_{\phi}$) is usually investigated, also sweeping both positive and negative values of current and magnetic field. Thus, we study the modulation of $I_c(H)$ for both the positive ($I^+_c > 0$) and negative ($I^-_c < 0$) directions of the bias current injected into the system. The pattern $I_c^\pm(H)$ can be characterized by three different types of symmetry: (\emph{i}) current inversion symmetry (CIS), $I^+_c (B) = -I^-_c (B)$, see Fig.~\ref{SuppMat5}a); (\emph{ii}) magnetic inversion symmetry (MIS), $I^\pm_c (B) = I^\pm_c (-B)$, see purple and red curves in Fig.~\ref{SuppMat5}b); (\emph{iii}) current and magnetic inversion symmetry (CMIS), $I^\pm_c (B) = -I^\mp_c (-B)$, see purple and blue curves in Fig.~\ref{SuppMat5}b).

A detailed overview of CPRs, the connections with the breaking of system's intrinsic symmetries, and the resulting symmetry of $I_c^\pm(H)$ patterns are thoroughly presented in Ref.~\cite{Kashiwaya19}. For our purposes, we recall that for a time-invariant system, CMIS is expected. Taking a cue from this observation, Singh \emph{et al.}~\cite{NPJ066} account for the observed critical current patterns of LAO/STO junctions, which were not symmetric under CI or MI, but showed a clear CMIS, i.e., $I^\pm_c (H) = -I^\mp_c (-H)$, through a minimal three-channel phenomenological model that does not include TRS-breaking contributions.

Now, taking the CPRs obtained from our microscopic approach, the symmetries underlying the $I_c^\pm$ patterns can be determined by the resistively shunted junction (RSJ) model~\cite{Barone82}, considering both positive/negative flowing directions of the bias current. Within this model, a current-biased JJ is depicted as the parallel between different current contributions: the external bias current, the Josephson term due to the Cooper-pair dissipationless flow (i.e., the CPR), the resistive contribution from the quasiparticle tunneling, the displacement current, and a thermal current. The latter two contributions are ignored in our simplified approach, which has the sole purpose of recovering the current/magnetic inversion symmetry of $I_c$. In fact, for the determination of this symmetry the occurrence of premature switching induced by thermal fluctuations and the increase in inertia of the phase particle are not relevant. Therefore, normalizing the time to the inverse of the plasma frequency~\cite{Barone82}, the RSJ model, in dimensionless units, can be easily written as $I(\phi)+d\phi/dt=I_{bias}(t)$, where we assumed a linearly ramping bias current $I_{bias}(t)=\pm t/t_{max}$, with $t_{max}$ being the preset measurement time. 
According to this model, the "phase particle" is confined within a minimum of the so-called tilted washboard potential; as the current increases, the height of the confining potential barrier gradually decreases, to the point where the phase particle can "escape" the potential minimum and roll along the potential profile. When this happens, a finite voltage $V \propto d\phi/dt$ appears. Therefore, by slowly ramping $I_{bias}$, we record the bias current value at which the system switches towards the voltage state. This "experiment" is finally repeated by reversing $I_{bias}$ and changing the value of $M_x$, the latter giving a specific CPR, as shown in Fig.~\ref{SuppMat5}c). The resulting switching current profiles, $I_c^\pm(M_x)$, are illustrated in Fig.~\ref{SuppMat5}d). The CMIS is clearly evident, just in line with the experimental findings on the unconventional critical current patterns of non-centrosymmetric materials, e.g., see Fig.~3 of Ref.~\cite{NPJ066}.
\section{Comparison between a magnetic field along x and z directions.}
\label{AppF}
\begin{figure}
	\includegraphics[scale=0.2]{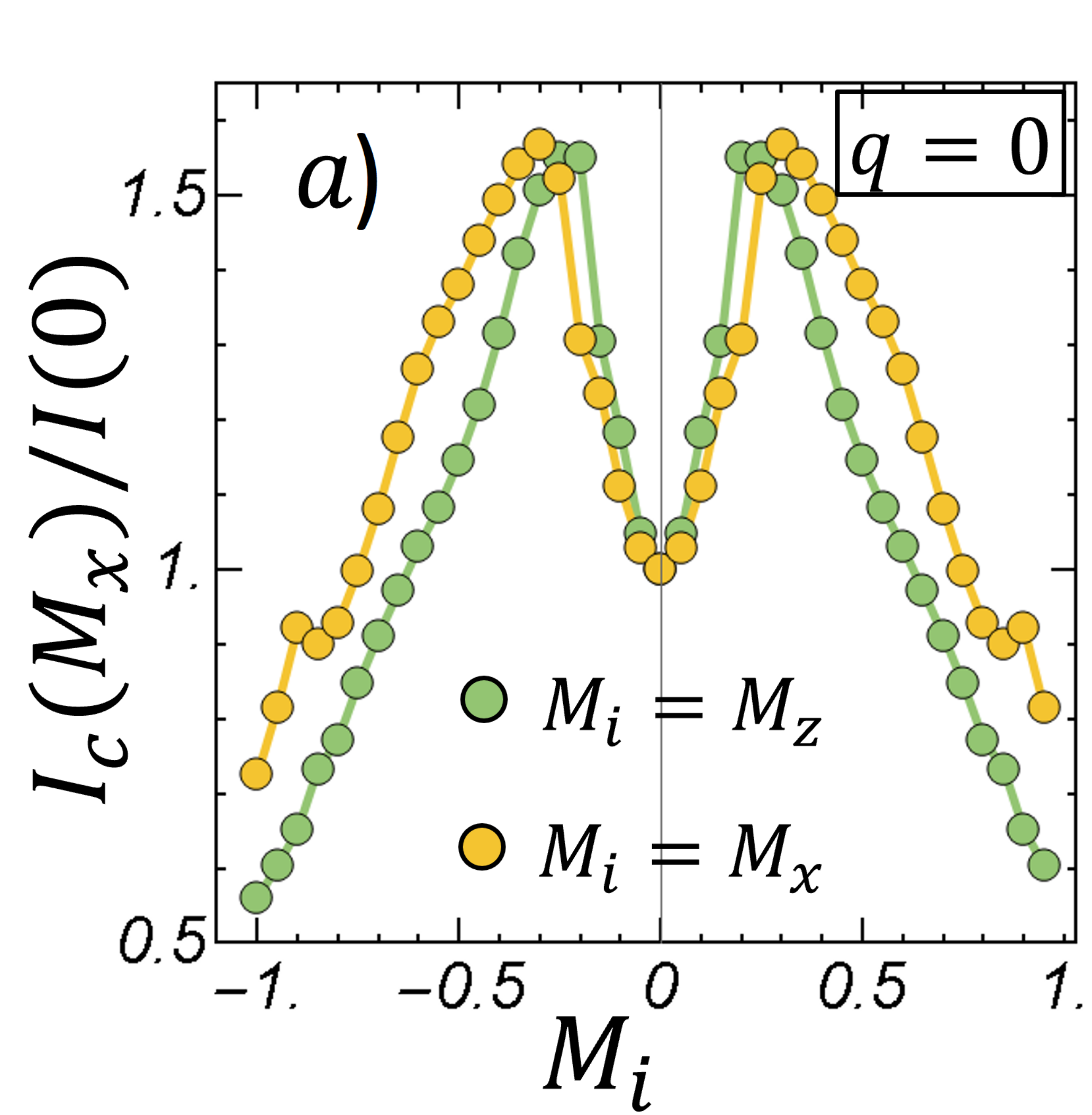}
	\includegraphics[scale=0.2]{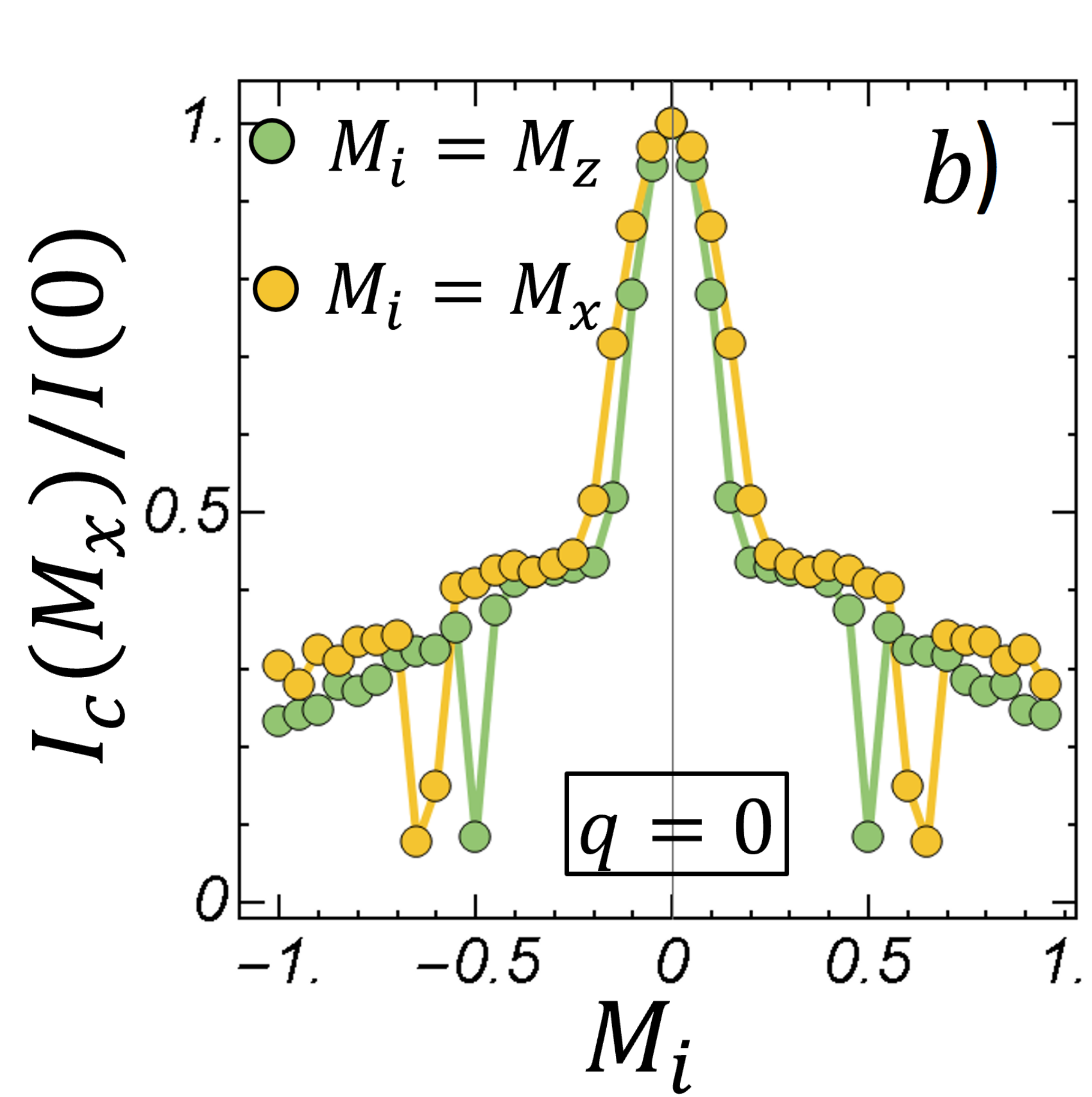}
	\caption{Comparison between $M_x$ and $M_z$ for the critical current curves when spin and momentum are locked, panel a), and unlocked, panel b). Quantitative discrepancies are manifested for $M_i \geq 0.3$ while the physical trend remains the same for the entire range.}
	\label{SuppMat4}
\end{figure}
For fillings belonging to orbital $d_{xy}$, the analysis of the main text has been performed by fixing $M_y=M_z=0$ and $M_x \neq 0$. However, the key ingredient for achieving a topological phase is to consider a magnetic field perpendicular to the orientation of the orbital Rashba field ($\sigma_y$) \cite{PhysRevB.100.094526}. In this section we compare some $I_c$ curves of the main text with those obtained by  $M_y=M_x=0$ and $M_z \neq 0$. We observe, in Fig.~\ref{SuppMat4}, a semi-quantitative agreement between the two orientations of magnetic fields, both perpendicular to $\emph{y}$-direction. The agreement persists both in a topological point of the phase diagram, panel a), and also when the junction experiences a trivial regime, panel b). However, for $M_i \geq 0.3$ with $i=x,y$, some quantitative discrepancies are clearly observed due to the multiorbital nature of the described nanochannel. Indeed, it is straightforward to show that for a nanowire with one single orbital, spin-orbit and magnetic terms, a unitary matrix transforms $M_x$ into $M_y$ without affecting the remaining terms of the Hamiltonian.
\bibliographystyle{apsrev}
\bibliography{Bib}

\begin{thebibliography}{53}
\expandafter\ifx\csname natexlab\endcsname\relax\def\natexlab#1{#1}\fi
\expandafter\ifx\csname bibnamefont\endcsname\relax
  \def\bibnamefont#1{#1}\fi
\expandafter\ifx\csname bibfnamefont\endcsname\relax
  \def\bibfnamefont#1{#1}\fi
\expandafter\ifx\csname citenamefont\endcsname\relax
  \def\citenamefont#1{#1}\fi
\expandafter\ifx\csname url\endcsname\relax
  \def\url#1{\texttt{#1}}\fi
\expandafter\ifx\csname urlprefix\endcsname\relax\def\urlprefix{URL }\fi
\providecommand{\bibinfo}[2]{#2}
\providecommand{\eprint}[2][]{\url{#2}}

\bibitem[{\citenamefont{Lutchyn et~al.}(2010)\citenamefont{Lutchyn, Sau, and
  Das~Sarma}}]{Lutchyn_2010}
\bibinfo{author}{\bibfnamefont{R.~M.} \bibnamefont{Lutchyn}},
  \bibinfo{author}{\bibfnamefont{J.~D.} \bibnamefont{Sau}}, \bibnamefont{and}
  \bibinfo{author}{\bibfnamefont{S.}~\bibnamefont{Das~Sarma}},
  \bibinfo{journal}{Phys. Rev. Lett.} \textbf{\bibinfo{volume}{105}},
  \bibinfo{pages}{077001} (\bibinfo{year}{2010}),
  \urlprefix\url{https://link.aps.org/doi/10.1103/PhysRevLett.105.077001}.

\bibitem[{\citenamefont{Maiellaro et~al.}(2019)\citenamefont{Maiellaro, Romeo,
  Perroni, Cataudella, and Citro}}]{Maiellaro2019}
\bibinfo{author}{\bibfnamefont{A.}~\bibnamefont{Maiellaro}},
  \bibinfo{author}{\bibfnamefont{F.}~\bibnamefont{Romeo}},
  \bibinfo{author}{\bibfnamefont{C.~A.} \bibnamefont{Perroni}},
  \bibinfo{author}{\bibfnamefont{V.}~\bibnamefont{Cataudella}},
  \bibnamefont{and} \bibinfo{author}{\bibfnamefont{R.}~\bibnamefont{Citro}},
  \bibinfo{journal}{Nanomaterials} \textbf{\bibinfo{volume}{9}},
  \bibinfo{pages}{894} (\bibinfo{year}{2019}),
  \urlprefix\url{https://www.mdpi.com/2079-4991/9/6/894}.

\bibitem[{\citenamefont{Oreg et~al.}(2010)\citenamefont{Oreg, Refael, and von
  Oppen}}]{Oreg_2010}
\bibinfo{author}{\bibfnamefont{Y.}~\bibnamefont{Oreg}},
  \bibinfo{author}{\bibfnamefont{G.}~\bibnamefont{Refael}}, \bibnamefont{and}
  \bibinfo{author}{\bibfnamefont{F.}~\bibnamefont{von Oppen}},
  \bibinfo{journal}{Phys. Rev. Lett.} \textbf{\bibinfo{volume}{105}},
  \bibinfo{pages}{177002} (\bibinfo{year}{2010}),
  \urlprefix\url{https://link.aps.org/doi/10.1103/PhysRevLett.105.177002}.

\bibitem[{\citenamefont{Sau et~al.}(2010)\citenamefont{Sau, Lutchyn, Tewari,
  and Das~Sarma}}]{sau_2010}
\bibinfo{author}{\bibfnamefont{J.~D.} \bibnamefont{Sau}},
  \bibinfo{author}{\bibfnamefont{R.~M.} \bibnamefont{Lutchyn}},
  \bibinfo{author}{\bibfnamefont{S.}~\bibnamefont{Tewari}}, \bibnamefont{and}
  \bibinfo{author}{\bibfnamefont{S.}~\bibnamefont{Das~Sarma}},
  \bibinfo{journal}{Phys. Rev. Lett.} \textbf{\bibinfo{volume}{104}},
  \bibinfo{pages}{040502} (\bibinfo{year}{2010}),
  \urlprefix\url{https://link.aps.org/doi/10.1103/PhysRevLett.104.040502}.

\bibitem[{\citenamefont{Aguado}(2017)}]{Aguado_2017}
\bibinfo{author}{\bibfnamefont{R.}~\bibnamefont{Aguado}},
  \bibinfo{journal}{Riv. Nuovo Cim.} \textbf{\bibinfo{volume}{40}},
  \bibinfo{pages}{523} (\bibinfo{year}{2017}),
  \eprint{10.1393/ncr/i2017-10141-9},
  \urlprefix\url{https://www.sif.it/riviste/sif/ncr/econtents/2017/040/11/article/0}.

\bibitem[{\citenamefont{Mourik et~al.}(2012)\citenamefont{Mourik, Zuo, Frolov,
  Plissard, Bakkers, and Kouwenhoven}}]{doi:10.1126/science.1222360}
\bibinfo{author}{\bibfnamefont{V.}~\bibnamefont{Mourik}},
  \bibinfo{author}{\bibfnamefont{K.}~\bibnamefont{Zuo}},
  \bibinfo{author}{\bibfnamefont{S.~M.} \bibnamefont{Frolov}},
  \bibinfo{author}{\bibfnamefont{S.~R.} \bibnamefont{Plissard}},
  \bibinfo{author}{\bibfnamefont{E.~P. A.~M.} \bibnamefont{Bakkers}},
  \bibnamefont{and} \bibinfo{author}{\bibfnamefont{L.~P.}
  \bibnamefont{Kouwenhoven}}, \bibinfo{journal}{Science}
  \textbf{\bibinfo{volume}{336}}, \bibinfo{pages}{1003} (\bibinfo{year}{2012}),
  \urlprefix\url{https://www.science.org/doi/abs/10.1126/science.1222360}.

\bibitem[{\citenamefont{Das et~al.}(2012)\citenamefont{Das, Ronen, Most, and
  et~al.}}]{Nat10.1038}
\bibinfo{author}{\bibfnamefont{A.}~\bibnamefont{Das}},
  \bibinfo{author}{\bibfnamefont{Y.}~\bibnamefont{Ronen}},
  \bibinfo{author}{\bibfnamefont{Y.}~\bibnamefont{Most}}, \bibnamefont{and}
  \bibinfo{author}{\bibnamefont{et~al.}}, \bibinfo{journal}{Nature Phys}
  \textbf{\bibinfo{volume}{8}}, \bibinfo{pages}{887} (\bibinfo{year}{2012}),
  \urlprefix\url{https://doi.org/10.1038/nphys2479}.

\bibitem[{\citenamefont{Deng et~al.}(2014)\citenamefont{Deng, Yu, Huang, and
  et~al.}}]{Scirep10.1038}
\bibinfo{author}{\bibfnamefont{M.}~\bibnamefont{Deng}},
  \bibinfo{author}{\bibfnamefont{C.}~\bibnamefont{Yu}},
  \bibinfo{author}{\bibfnamefont{G.}~\bibnamefont{Huang}}, \bibnamefont{and}
  \bibinfo{author}{\bibnamefont{et~al.}}, \bibinfo{journal}{Sci Rep}
  \textbf{\bibinfo{volume}{4}}, \bibinfo{pages}{7261} (\bibinfo{year}{2014}),
  \urlprefix\url{https://doi.org/10.1038/srep07261}.

\bibitem[{\citenamefont{Albrecht et~al.}(2016)\citenamefont{Albrecht,
  Higginbotham, and et~al.}}]{Nat17162}
\bibinfo{author}{\bibfnamefont{S.}~\bibnamefont{Albrecht}},
  \bibinfo{author}{\bibfnamefont{M.}~\bibnamefont{Higginbotham},
  \bibfnamefont{A.and~Madsen}}, \bibnamefont{and}
  \bibinfo{author}{\bibnamefont{et~al.}}, \bibinfo{journal}{Nature}
  \textbf{\bibinfo{volume}{531}}, \bibinfo{pages}{206} (\bibinfo{year}{2016}),
  \urlprefix\url{https://doi.org/10.1038/nature17162}.

\bibitem[{\citenamefont{Zhang et~al.}(2018{\natexlab{a}})\citenamefont{Zhang,
  Liu, Gazibegovic, and et~al.}}]{zhang_2018}
\bibinfo{author}{\bibfnamefont{H.}~\bibnamefont{Zhang}},
  \bibinfo{author}{\bibfnamefont{C.}~\bibnamefont{Liu}},
  \bibinfo{author}{\bibfnamefont{S.}~\bibnamefont{Gazibegovic}},
  \bibnamefont{and} \bibinfo{author}{\bibnamefont{et~al.}},
  \bibinfo{journal}{Nature} \textbf{\bibinfo{volume}{556}}, \bibinfo{pages}{74}
  (\bibinfo{year}{2018}{\natexlab{a}}),
  \urlprefix\url{https://doi.org/10.1038/nature26142}.

\bibitem[{\citenamefont{Maiellaro and Citro}(2021)}]{condmat6020015}
\bibinfo{author}{\bibfnamefont{A.}~\bibnamefont{Maiellaro}} \bibnamefont{and}
  \bibinfo{author}{\bibfnamefont{R.}~\bibnamefont{Citro}},
  \bibinfo{journal}{Condensed Matter} \textbf{\bibinfo{volume}{6}}
  (\bibinfo{year}{2021}), ISSN \bibinfo{issn}{2410-3896},
  \urlprefix\url{https://www.mdpi.com/2410-3896/6/2/15}.

\bibitem[{\citenamefont{Zhang et~al.}(2018{\natexlab{b}})\citenamefont{Zhang,
  Yaji, Hashimoto, Ota, Kondo, Okazaki, Wang, Wen, Gu, Ding
  et~al.}}]{zhang_iron_2018}
\bibinfo{author}{\bibfnamefont{P.}~\bibnamefont{Zhang}},
  \bibinfo{author}{\bibfnamefont{K.}~\bibnamefont{Yaji}},
  \bibinfo{author}{\bibfnamefont{T.}~\bibnamefont{Hashimoto}},
  \bibinfo{author}{\bibfnamefont{Y.}~\bibnamefont{Ota}},
  \bibinfo{author}{\bibfnamefont{T.}~\bibnamefont{Kondo}},
  \bibinfo{author}{\bibfnamefont{K.}~\bibnamefont{Okazaki}},
  \bibinfo{author}{\bibfnamefont{Z.}~\bibnamefont{Wang}},
  \bibinfo{author}{\bibfnamefont{J.}~\bibnamefont{Wen}},
  \bibinfo{author}{\bibfnamefont{G.~D.} \bibnamefont{Gu}},
  \bibinfo{author}{\bibfnamefont{H.}~\bibnamefont{Ding}}, \bibnamefont{et~al.},
  \bibinfo{journal}{Science} \textbf{\bibinfo{volume}{360}},
  \bibinfo{pages}{182} (\bibinfo{year}{2018}{\natexlab{b}}),
  \urlprefix\url{https://www.science.org/doi/abs/10.1126/science.aan4596}.

\bibitem[{\citenamefont{Maiellaro et~al.}(2020)\citenamefont{Maiellaro, Romeo,
  and Citro}}]{MaiellaroEPJst}
\bibinfo{author}{\bibfnamefont{A.}~\bibnamefont{Maiellaro}},
  \bibinfo{author}{\bibfnamefont{F.}~\bibnamefont{Romeo}}, \bibnamefont{and}
  \bibinfo{author}{\bibfnamefont{R.}~\bibnamefont{Citro}},
  \bibinfo{journal}{Eur. Phys. J. Spec. Top.} \textbf{\bibinfo{volume}{229}},
  \bibinfo{pages}{637} (\bibinfo{year}{2020}),
  \urlprefix\url{https://doi.org/10.1140/epjst/e2019-900180-x}.

\bibitem[{\citenamefont{Wray et~al.}(2011)\citenamefont{Wray, Xu, Xia, Hsieh,
  Fedorov, Hor, Cava, Bansil, Lin, and Hasan}}]{wray_2011}
\bibinfo{author}{\bibfnamefont{L.~A.} \bibnamefont{Wray}},
  \bibinfo{author}{\bibfnamefont{S.-Y.} \bibnamefont{Xu}},
  \bibinfo{author}{\bibfnamefont{Y.}~\bibnamefont{Xia}},
  \bibinfo{author}{\bibfnamefont{D.}~\bibnamefont{Hsieh}},
  \bibinfo{author}{\bibfnamefont{A.~V.} \bibnamefont{Fedorov}},
  \bibinfo{author}{\bibfnamefont{Y.~S.} \bibnamefont{Hor}},
  \bibinfo{author}{\bibfnamefont{R.~J.} \bibnamefont{Cava}},
  \bibinfo{author}{\bibfnamefont{A.}~\bibnamefont{Bansil}},
  \bibinfo{author}{\bibfnamefont{H.}~\bibnamefont{Lin}}, \bibnamefont{and}
  \bibinfo{author}{\bibfnamefont{M.~Z.} \bibnamefont{Hasan}},
  \bibinfo{journal}{Nature Physics} \textbf{\bibinfo{volume}{7}},
  \bibinfo{pages}{32} (\bibinfo{year}{2011}),
  \urlprefix\url{https://doi.org/10.1038/nphys1838}.

\bibitem[{\citenamefont{Fornieri et~al.}(2019)\citenamefont{Fornieri, Whiticar,
  Setiawan, Portol{\'e}s, Drachmann, Keselman, Gronin, Thomas, Wang, Kallaher
  et~al.}}]{fornieri_2019}
\bibinfo{author}{\bibfnamefont{A.}~\bibnamefont{Fornieri}},
  \bibinfo{author}{\bibfnamefont{A.~M.} \bibnamefont{Whiticar}},
  \bibinfo{author}{\bibfnamefont{F.}~\bibnamefont{Setiawan}},
  \bibinfo{author}{\bibfnamefont{E.}~\bibnamefont{Portol{\'e}s}},
  \bibinfo{author}{\bibfnamefont{A.~C.} \bibnamefont{Drachmann}},
  \bibinfo{author}{\bibfnamefont{A.}~\bibnamefont{Keselman}},
  \bibinfo{author}{\bibfnamefont{S.}~\bibnamefont{Gronin}},
  \bibinfo{author}{\bibfnamefont{C.}~\bibnamefont{Thomas}},
  \bibinfo{author}{\bibfnamefont{T.}~\bibnamefont{Wang}},
  \bibinfo{author}{\bibfnamefont{R.}~\bibnamefont{Kallaher}},
  \bibnamefont{et~al.}, \bibinfo{journal}{Nature}
  \textbf{\bibinfo{volume}{569}}, \bibinfo{pages}{89} (\bibinfo{year}{2019}),
  \urlprefix\url{https://doi.org/10.1038/s41586-019-1068-8}.

\bibitem[{\citenamefont{Maiellaro et~al.}(2022)\citenamefont{Maiellaro,
  Illuminati, and Citro}}]{condmat7010026}
\bibinfo{author}{\bibfnamefont{A.}~\bibnamefont{Maiellaro}},
  \bibinfo{author}{\bibfnamefont{F.}~\bibnamefont{Illuminati}},
  \bibnamefont{and} \bibinfo{author}{\bibfnamefont{R.}~\bibnamefont{Citro}},
  \bibinfo{journal}{Condensed Matter} \textbf{\bibinfo{volume}{7}}
  (\bibinfo{year}{2022}), ISSN \bibinfo{issn}{2410-3896},
  \urlprefix\url{https://www.mdpi.com/2410-3896/7/1/26}.

\bibitem[{\citenamefont{Barthelemy et~al.}(2021)\citenamefont{Barthelemy,
  Bergeal, Bibes, Caviglia, Citro, Cuoco, Kalaboukhov, Kalisky, Perroni,
  Santamaria et~al.}}]{barthelemy_2021}
\bibinfo{author}{\bibfnamefont{A.}~\bibnamefont{Barthelemy}},
  \bibinfo{author}{\bibfnamefont{N.}~\bibnamefont{Bergeal}},
  \bibinfo{author}{\bibfnamefont{M.}~\bibnamefont{Bibes}},
  \bibinfo{author}{\bibfnamefont{A.}~\bibnamefont{Caviglia}},
  \bibinfo{author}{\bibfnamefont{R.}~\bibnamefont{Citro}},
  \bibinfo{author}{\bibfnamefont{M.}~\bibnamefont{Cuoco}},
  \bibinfo{author}{\bibfnamefont{A.}~\bibnamefont{Kalaboukhov}},
  \bibinfo{author}{\bibfnamefont{B.}~\bibnamefont{Kalisky}},
  \bibinfo{author}{\bibfnamefont{C.~A.} \bibnamefont{Perroni}},
  \bibinfo{author}{\bibfnamefont{J.}~\bibnamefont{Santamaria}},
  \bibnamefont{et~al.}, \bibinfo{journal}{Europhysics Letters}
  \textbf{\bibinfo{volume}{133}}, \bibinfo{pages}{17001}
  (\bibinfo{year}{2021}),
  \urlprefix\url{https://dx.doi.org/10.1209/0295-5075/133/17001}.

\bibitem[{\citenamefont{Settino et~al.}(2021)\citenamefont{Settino, Citro,
  Romeo, Cataudella, and Perroni}}]{PhysRevB.103.235120}
\bibinfo{author}{\bibfnamefont{J.}~\bibnamefont{Settino}},
  \bibinfo{author}{\bibfnamefont{R.}~\bibnamefont{Citro}},
  \bibinfo{author}{\bibfnamefont{F.}~\bibnamefont{Romeo}},
  \bibinfo{author}{\bibfnamefont{V.}~\bibnamefont{Cataudella}},
  \bibnamefont{and} \bibinfo{author}{\bibfnamefont{C.~A.}
  \bibnamefont{Perroni}}, \bibinfo{journal}{Phys. Rev. B}
  \textbf{\bibinfo{volume}{103}}, \bibinfo{pages}{235120}
  (\bibinfo{year}{2021}),
  \urlprefix\url{https://link.aps.org/doi/10.1103/PhysRevB.103.235120}.

\bibitem[{\citenamefont{Scheurer and Schmalian}(2015)}]{schmalian_2015}
\bibinfo{author}{\bibfnamefont{M.~S.} \bibnamefont{Scheurer}} \bibnamefont{and}
  \bibinfo{author}{\bibfnamefont{J.}~\bibnamefont{Schmalian}},
  \bibinfo{journal}{Nature communications} \textbf{\bibinfo{volume}{6}},
  \bibinfo{pages}{1} (\bibinfo{year}{2015}),
  \urlprefix\url{https://doi.org/10.1038/ncomms7005}.

\bibitem[{\citenamefont{Mazziotti et~al.}(2018)\citenamefont{Mazziotti,
  Scopigno, Grilli, and Caprara}}]{mazziotti_2018}
\bibinfo{author}{\bibfnamefont{M.~V.} \bibnamefont{Mazziotti}},
  \bibinfo{author}{\bibfnamefont{N.}~\bibnamefont{Scopigno}},
  \bibinfo{author}{\bibfnamefont{M.}~\bibnamefont{Grilli}}, \bibnamefont{and}
  \bibinfo{author}{\bibfnamefont{S.}~\bibnamefont{Caprara}},
  \bibinfo{journal}{Condensed Matter} \textbf{\bibinfo{volume}{3}}
  (\bibinfo{year}{2018}), ISSN \bibinfo{issn}{2410-3896},
  \urlprefix\url{https://www.mdpi.com/2410-3896/3/4/37}.

\bibitem[{\citenamefont{Perroni et~al.}(2019)\citenamefont{Perroni, Cataudella,
  Salluzzo, Cuoco, and Citro}}]{PhysRevB.100.094526}
\bibinfo{author}{\bibfnamefont{C.~A.} \bibnamefont{Perroni}},
  \bibinfo{author}{\bibfnamefont{V.}~\bibnamefont{Cataudella}},
  \bibinfo{author}{\bibfnamefont{M.}~\bibnamefont{Salluzzo}},
  \bibinfo{author}{\bibfnamefont{M.}~\bibnamefont{Cuoco}}, \bibnamefont{and}
  \bibinfo{author}{\bibfnamefont{R.}~\bibnamefont{Citro}},
  \bibinfo{journal}{Phys. Rev. B} \textbf{\bibinfo{volume}{100}},
  \bibinfo{pages}{094526} (\bibinfo{year}{2019}),
  \urlprefix\url{https://link.aps.org/doi/10.1103/PhysRevB.100.094526}.

\bibitem[{\citenamefont{Caviglia et~al.}(2010)\citenamefont{Caviglia, Gabay,
  Gariglio, Reyren, Cancellieri, and Triscone}}]{caviglia_2010}
\bibinfo{author}{\bibfnamefont{A.~D.} \bibnamefont{Caviglia}},
  \bibinfo{author}{\bibfnamefont{M.}~\bibnamefont{Gabay}},
  \bibinfo{author}{\bibfnamefont{S.}~\bibnamefont{Gariglio}},
  \bibinfo{author}{\bibfnamefont{N.}~\bibnamefont{Reyren}},
  \bibinfo{author}{\bibfnamefont{C.}~\bibnamefont{Cancellieri}},
  \bibnamefont{and} \bibinfo{author}{\bibfnamefont{J.-M.}
  \bibnamefont{Triscone}}, \bibinfo{journal}{Phys. Rev. Lett.}
  \textbf{\bibinfo{volume}{104}}, \bibinfo{pages}{126803}
  (\bibinfo{year}{2010}),
  \urlprefix\url{https://link.aps.org/doi/10.1103/PhysRevLett.104.126803}.

\bibitem[{\citenamefont{Reyren et~al.}(2007)\citenamefont{Reyren, Thiel,
  Caviglia, Kourkoutis, Hammerl, Richter, Schneider, Kopp, Rüetschi, Jaccard
  et~al.}}]{reyren_2007}
\bibinfo{author}{\bibfnamefont{N.}~\bibnamefont{Reyren}},
  \bibinfo{author}{\bibfnamefont{S.}~\bibnamefont{Thiel}},
  \bibinfo{author}{\bibfnamefont{A.~D.} \bibnamefont{Caviglia}},
  \bibinfo{author}{\bibfnamefont{L.~F.} \bibnamefont{Kourkoutis}},
  \bibinfo{author}{\bibfnamefont{G.}~\bibnamefont{Hammerl}},
  \bibinfo{author}{\bibfnamefont{C.}~\bibnamefont{Richter}},
  \bibinfo{author}{\bibfnamefont{C.~W.} \bibnamefont{Schneider}},
  \bibinfo{author}{\bibfnamefont{T.}~\bibnamefont{Kopp}},
  \bibinfo{author}{\bibfnamefont{A.-S.} \bibnamefont{Rüetschi}},
  \bibinfo{author}{\bibfnamefont{D.}~\bibnamefont{Jaccard}},
  \bibnamefont{et~al.}, \bibinfo{journal}{Science}
  \textbf{\bibinfo{volume}{317}}, \bibinfo{pages}{1196} (\bibinfo{year}{2007}),
  \urlprefix\url{https://www.science.org/doi/abs/10.1126/science.1146006}.

\bibitem[{\citenamefont{Caviglia et~al.}(2008)\citenamefont{Caviglia, Gariglio,
  Reyren, and et~al.}}]{caviglia_2008}
\bibinfo{author}{\bibfnamefont{A.}~\bibnamefont{Caviglia}},
  \bibinfo{author}{\bibfnamefont{S.}~\bibnamefont{Gariglio}},
  \bibinfo{author}{\bibfnamefont{N.}~\bibnamefont{Reyren}}, \bibnamefont{and}
  \bibinfo{author}{\bibnamefont{et~al.}}, \bibinfo{journal}{Nature}
  \textbf{\bibinfo{volume}{456}}, \bibinfo{pages}{624} (\bibinfo{year}{2008}),
  \urlprefix\url{https://doi.org/10.1038/nature07576}.

\bibitem[{\citenamefont{Salluzzo et~al.}(2009)\citenamefont{Salluzzo, Cezar,
  Brookes, Bisogni, De~Luca, Richter, Thiel, Mannhart, Huijben, Brinkman
  et~al.}}]{salluzzo_2009}
\bibinfo{author}{\bibfnamefont{M.}~\bibnamefont{Salluzzo}},
  \bibinfo{author}{\bibfnamefont{J.~C.} \bibnamefont{Cezar}},
  \bibinfo{author}{\bibfnamefont{N.~B.} \bibnamefont{Brookes}},
  \bibinfo{author}{\bibfnamefont{V.}~\bibnamefont{Bisogni}},
  \bibinfo{author}{\bibfnamefont{G.~M.} \bibnamefont{De~Luca}},
  \bibinfo{author}{\bibfnamefont{C.}~\bibnamefont{Richter}},
  \bibinfo{author}{\bibfnamefont{S.}~\bibnamefont{Thiel}},
  \bibinfo{author}{\bibfnamefont{J.}~\bibnamefont{Mannhart}},
  \bibinfo{author}{\bibfnamefont{M.}~\bibnamefont{Huijben}},
  \bibinfo{author}{\bibfnamefont{A.}~\bibnamefont{Brinkman}},
  \bibnamefont{et~al.}, \bibinfo{journal}{Phys. Rev. Lett.}
  \textbf{\bibinfo{volume}{102}}, \bibinfo{pages}{166804}
  (\bibinfo{year}{2009}),
  \urlprefix\url{https://link.aps.org/doi/10.1103/PhysRevLett.102.166804}.

\bibitem[{\citenamefont{Stornaiuolo et~al.}(2017)\citenamefont{Stornaiuolo,
  Massarotti, Di~Capua, Lucignano, Pepe, Salluzzo, and
  Tafuri}}]{PhysRevB.95.140502}
\bibinfo{author}{\bibfnamefont{D.}~\bibnamefont{Stornaiuolo}},
  \bibinfo{author}{\bibfnamefont{D.}~\bibnamefont{Massarotti}},
  \bibinfo{author}{\bibfnamefont{R.}~\bibnamefont{Di~Capua}},
  \bibinfo{author}{\bibfnamefont{P.}~\bibnamefont{Lucignano}},
  \bibinfo{author}{\bibfnamefont{G.~P.} \bibnamefont{Pepe}},
  \bibinfo{author}{\bibfnamefont{M.}~\bibnamefont{Salluzzo}}, \bibnamefont{and}
  \bibinfo{author}{\bibfnamefont{F.}~\bibnamefont{Tafuri}},
  \bibinfo{journal}{Phys. Rev. B} \textbf{\bibinfo{volume}{95}},
  \bibinfo{pages}{140502} (\bibinfo{year}{2017}),
  \urlprefix\url{https://link.aps.org/doi/10.1103/PhysRevB.95.140502}.

\bibitem[{\citenamefont{Singh et~al.}(2022)\citenamefont{Singh, Guarcello,
  Lesne, and et~al.}}]{NPJ066}
\bibinfo{author}{\bibfnamefont{G.}~\bibnamefont{Singh}},
  \bibinfo{author}{\bibfnamefont{C.}~\bibnamefont{Guarcello}},
  \bibinfo{author}{\bibfnamefont{E.}~\bibnamefont{Lesne}}, \bibnamefont{and}
  \bibinfo{author}{\bibnamefont{et~al.}}, \bibinfo{journal}{npj Quantum Matter}
  \textbf{\bibinfo{volume}{7}}, \bibinfo{pages}{2} (\bibinfo{year}{2022}),
  \urlprefix\url{https://doi.org/10.1038/s41535-021-00406-6}.

\bibitem[{\citenamefont{Fukaya et~al.}(2018)\citenamefont{Fukaya, Tamura, Yada,
  Tanaka, Gentile, and Cuoco}}]{PhysRevB.97.174522}
\bibinfo{author}{\bibfnamefont{Y.}~\bibnamefont{Fukaya}},
  \bibinfo{author}{\bibfnamefont{S.}~\bibnamefont{Tamura}},
  \bibinfo{author}{\bibfnamefont{K.}~\bibnamefont{Yada}},
  \bibinfo{author}{\bibfnamefont{Y.}~\bibnamefont{Tanaka}},
  \bibinfo{author}{\bibfnamefont{P.}~\bibnamefont{Gentile}}, \bibnamefont{and}
  \bibinfo{author}{\bibfnamefont{M.}~\bibnamefont{Cuoco}},
  \bibinfo{journal}{Phys. Rev. B} \textbf{\bibinfo{volume}{97}},
  \bibinfo{pages}{174522} (\bibinfo{year}{2018}),
  \urlprefix\url{https://link.aps.org/doi/10.1103/PhysRevB.97.174522}.

\bibitem[{\citenamefont{Tiira et~al.}(2017)\citenamefont{Tiira, Strambini,
  Amado, and et~al.}}]{NatCommun14984}
\bibinfo{author}{\bibfnamefont{J.}~\bibnamefont{Tiira}},
  \bibinfo{author}{\bibfnamefont{E.}~\bibnamefont{Strambini}},
  \bibinfo{author}{\bibfnamefont{M.}~\bibnamefont{Amado}}, \bibnamefont{and}
  \bibinfo{author}{\bibnamefont{et~al.}}, \bibinfo{journal}{Nat Commun}
  \textbf{\bibinfo{volume}{8}}, \bibinfo{pages}{14984} (\bibinfo{year}{2017}),
  \urlprefix\url{https://doi.org/10.1038/ncomms14984}.

\bibitem[{\citenamefont{Settino et~al.}(2020)\citenamefont{Settino, Forte,
  Perroni, Cataudella, Cuoco, and Citro}}]{PhysRevB.102.224508}
\bibinfo{author}{\bibfnamefont{J.}~\bibnamefont{Settino}},
  \bibinfo{author}{\bibfnamefont{F.}~\bibnamefont{Forte}},
  \bibinfo{author}{\bibfnamefont{C.~A.} \bibnamefont{Perroni}},
  \bibinfo{author}{\bibfnamefont{V.}~\bibnamefont{Cataudella}},
  \bibinfo{author}{\bibfnamefont{M.}~\bibnamefont{Cuoco}}, \bibnamefont{and}
  \bibinfo{author}{\bibfnamefont{R.}~\bibnamefont{Citro}},
  \bibinfo{journal}{Phys. Rev. B} \textbf{\bibinfo{volume}{102}},
  \bibinfo{pages}{224508} (\bibinfo{year}{2020}),
  \urlprefix\url{https://link.aps.org/doi/10.1103/PhysRevB.102.224508}.

\bibitem[{\citenamefont{Pai et~al.}(2018)\citenamefont{Pai, Tylan-Tyler, Irvin,
  and Levy}}]{Pai_2018}
\bibinfo{author}{\bibfnamefont{Y.-Y.} \bibnamefont{Pai}},
  \bibinfo{author}{\bibfnamefont{A.}~\bibnamefont{Tylan-Tyler}},
  \bibinfo{author}{\bibfnamefont{P.}~\bibnamefont{Irvin}}, \bibnamefont{and}
  \bibinfo{author}{\bibfnamefont{J.}~\bibnamefont{Levy}},
  \bibinfo{journal}{Reports on Progress in Physics}
  \textbf{\bibinfo{volume}{81}}, \bibinfo{pages}{036503}
  (\bibinfo{year}{2018}),
  \urlprefix\url{https://doi.org/10.1088/1361-6633/aa892d}.

\bibitem[{\citenamefont{Zhong et~al.}(2013)\citenamefont{Zhong, T\'oth, and
  Held}}]{PhysRevB.87.161102}
\bibinfo{author}{\bibfnamefont{Z.}~\bibnamefont{Zhong}},
  \bibinfo{author}{\bibfnamefont{A.}~\bibnamefont{T\'oth}}, \bibnamefont{and}
  \bibinfo{author}{\bibfnamefont{K.}~\bibnamefont{Held}},
  \bibinfo{journal}{Phys. Rev. B} \textbf{\bibinfo{volume}{87}},
  \bibinfo{pages}{161102} (\bibinfo{year}{2013}),
  \urlprefix\url{https://link.aps.org/doi/10.1103/PhysRevB.87.161102}.

\bibitem[{\citenamefont{Joshua et~al.}(2012)\citenamefont{Joshua, Pecker,
  Ruhman, and et~al.}}]{Joshua12}
\bibinfo{author}{\bibfnamefont{A.}~\bibnamefont{Joshua}},
  \bibinfo{author}{\bibfnamefont{S.}~\bibnamefont{Pecker}},
  \bibinfo{author}{\bibfnamefont{J.}~\bibnamefont{Ruhman}}, \bibnamefont{and}
  \bibinfo{author}{\bibnamefont{et~al.}}, \bibinfo{journal}{Nat Commun}
  \textbf{\bibinfo{volume}{3}} (\bibinfo{year}{2012}),
  \urlprefix\url{https://doi.org/10.1038/ncomms2116}.

\bibitem[{\citenamefont{Yuan and Fu}(2022)}]{doi:10.1073/pnas.2119548119}
\bibinfo{author}{\bibfnamefont{N.~F.~Q.} \bibnamefont{Yuan}} \bibnamefont{and}
  \bibinfo{author}{\bibfnamefont{L.}~\bibnamefont{Fu}},
  \bibinfo{journal}{Proceedings of the National Academy of Sciences}
  \textbf{\bibinfo{volume}{119}}, \bibinfo{pages}{e2119548119}
  (\bibinfo{year}{2022}),
  \eprint{https://www.pnas.org/doi/pdf/10.1073/pnas.2119548119},
  \urlprefix\url{https://www.pnas.org/doi/abs/10.1073/pnas.2119548119}.

\bibitem[{\citenamefont{Cayao et~al.}(2017)\citenamefont{Cayao, San-Jose,
  Black-Schaffer, Aguado, and Prada}}]{PhysRevB.96.205425}
\bibinfo{author}{\bibfnamefont{J.}~\bibnamefont{Cayao}},
  \bibinfo{author}{\bibfnamefont{P.}~\bibnamefont{San-Jose}},
  \bibinfo{author}{\bibfnamefont{A.~M.} \bibnamefont{Black-Schaffer}},
  \bibinfo{author}{\bibfnamefont{R.}~\bibnamefont{Aguado}}, \bibnamefont{and}
  \bibinfo{author}{\bibfnamefont{E.}~\bibnamefont{Prada}},
  \bibinfo{journal}{Phys. Rev. B} \textbf{\bibinfo{volume}{96}},
  \bibinfo{pages}{205425} (\bibinfo{year}{2017}),
  \urlprefix\url{https://link.aps.org/doi/10.1103/PhysRevB.96.205425}.

\bibitem[{\citenamefont{Groth et~al.}(2014)\citenamefont{Groth, Wimmer,
  Akhmerov, and Waintal}}]{Groth_2014}
\bibinfo{author}{\bibfnamefont{C.~W.} \bibnamefont{Groth}},
  \bibinfo{author}{\bibfnamefont{M.}~\bibnamefont{Wimmer}},
  \bibinfo{author}{\bibfnamefont{A.~R.} \bibnamefont{Akhmerov}},
  \bibnamefont{and} \bibinfo{author}{\bibfnamefont{X.}~\bibnamefont{Waintal}},
  \bibinfo{journal}{New Journal of Physics} \textbf{\bibinfo{volume}{16}},
  \bibinfo{pages}{063065} (\bibinfo{year}{2014}),
  \urlprefix\url{https://doi.org/10.1088/1367-2630/16/6/063065}.

\bibitem[{\citenamefont{Oliphant}(2006)}]{citeulike:9919912}
\bibinfo{author}{\bibfnamefont{T.~E.} \bibnamefont{Oliphant}},
  \emph{\bibinfo{title}{{Guide to NumPy}}} (\bibinfo{publisher}{Trelgol},
  \bibinfo{year}{2006}),
  \urlprefix\url{https://www.bibsonomy.org/bibtex/f13bb8f9237574e5678cedd1b36c4175}.

\bibitem[{\citenamefont{Shen et~al.}(2016)\citenamefont{Shen, Chen, Xue, and
  et~al.}}]{Hc}
\bibinfo{author}{\bibfnamefont{S.}~\bibnamefont{Shen}},
  \bibinfo{author}{\bibfnamefont{B.}~\bibnamefont{Chen}},
  \bibinfo{author}{\bibfnamefont{H.}~\bibnamefont{Xue}}, \bibnamefont{and}
  \bibinfo{author}{\bibnamefont{et~al.}}, \bibinfo{journal}{Sci Rep}
  \textbf{\bibinfo{volume}{6}}, \bibinfo{pages}{28379} (\bibinfo{year}{2016}),
  \urlprefix\url{https://doi.org/10.1038/srep28379}.

\bibitem[{\citenamefont{Erlandsen et~al.}(2022)\citenamefont{Erlandsen, Dahm,
  Trier, Scuderi, Di~Gennaro, Sambri, Reffeldt~Kirchert, Pryds, Granozio, and
  Jespersen}}]{Hc2}
\bibinfo{author}{\bibfnamefont{R.}~\bibnamefont{Erlandsen}},
  \bibinfo{author}{\bibfnamefont{R.~T.} \bibnamefont{Dahm}},
  \bibinfo{author}{\bibfnamefont{F.}~\bibnamefont{Trier}},
  \bibinfo{author}{\bibfnamefont{M.}~\bibnamefont{Scuderi}},
  \bibinfo{author}{\bibfnamefont{E.}~\bibnamefont{Di~Gennaro}},
  \bibinfo{author}{\bibfnamefont{A.}~\bibnamefont{Sambri}},
  \bibinfo{author}{\bibfnamefont{C.~K.} \bibnamefont{Reffeldt~Kirchert}},
  \bibinfo{author}{\bibfnamefont{N.}~\bibnamefont{Pryds}},
  \bibinfo{author}{\bibfnamefont{F.~M.} \bibnamefont{Granozio}},
  \bibnamefont{and} \bibinfo{author}{\bibfnamefont{T.~S.}
  \bibnamefont{Jespersen}}, \bibinfo{journal}{Nano Letters}
  \textbf{\bibinfo{volume}{22}}, \bibinfo{pages}{4758} (\bibinfo{year}{2022}),
  \urlprefix\url{https://doi.org/10.1021/acs.nanolett.2c00992}.

\bibitem[{\citenamefont{Wang and Berggren}(1996)}]{PhysRevB.54.R14257}
\bibinfo{author}{\bibfnamefont{C.-K.} \bibnamefont{Wang}} \bibnamefont{and}
  \bibinfo{author}{\bibfnamefont{K.-F.} \bibnamefont{Berggren}},
  \bibinfo{journal}{Phys. Rev. B} \textbf{\bibinfo{volume}{54}},
  \bibinfo{pages}{R14257} (\bibinfo{year}{1996}),
  \urlprefix\url{https://link.aps.org/doi/10.1103/PhysRevB.54.R14257}.

\bibitem[{\citenamefont{Bj\o{}rlig et~al.}(2020)\citenamefont{Bj\o{}rlig,
  Carrad, Prawiroatmodjo, von Soosten, Gan, Chen, Pryds, Paaske, and
  Jespersen}}]{PhysRevMaterials.4.122001}
\bibinfo{author}{\bibfnamefont{A.~V.} \bibnamefont{Bj\o{}rlig}},
  \bibinfo{author}{\bibfnamefont{D.~J.} \bibnamefont{Carrad}},
  \bibinfo{author}{\bibfnamefont{G.~E. D.~K.} \bibnamefont{Prawiroatmodjo}},
  \bibinfo{author}{\bibfnamefont{M.}~\bibnamefont{von Soosten}},
  \bibinfo{author}{\bibfnamefont{Y.}~\bibnamefont{Gan}},
  \bibinfo{author}{\bibfnamefont{Y.}~\bibnamefont{Chen}},
  \bibinfo{author}{\bibfnamefont{N.}~\bibnamefont{Pryds}},
  \bibinfo{author}{\bibfnamefont{J.}~\bibnamefont{Paaske}}, \bibnamefont{and}
  \bibinfo{author}{\bibfnamefont{T.~S.} \bibnamefont{Jespersen}},
  \bibinfo{journal}{Phys. Rev. Materials} \textbf{\bibinfo{volume}{4}},
  \bibinfo{pages}{122001} (\bibinfo{year}{2020}),
  \urlprefix\url{https://link.aps.org/doi/10.1103/PhysRevMaterials.4.122001}.

\bibitem[{\citenamefont{Tian et~al.}(2014)\citenamefont{Tian, Childres, Cao,
  Shen, Miotkowski, and Chen}}]{TIAN20141}
\bibinfo{author}{\bibfnamefont{J.}~\bibnamefont{Tian}},
  \bibinfo{author}{\bibfnamefont{I.}~\bibnamefont{Childres}},
  \bibinfo{author}{\bibfnamefont{H.}~\bibnamefont{Cao}},
  \bibinfo{author}{\bibfnamefont{T.}~\bibnamefont{Shen}},
  \bibinfo{author}{\bibfnamefont{I.}~\bibnamefont{Miotkowski}},
  \bibnamefont{and} \bibinfo{author}{\bibfnamefont{Y.~P.} \bibnamefont{Chen}},
  \bibinfo{journal}{Solid State Communications} \textbf{\bibinfo{volume}{191}},
  \bibinfo{pages}{1} (\bibinfo{year}{2014}), ISSN \bibinfo{issn}{0038-1098},
  \urlprefix\url{https://www.sciencedirect.com/science/article/pii/S0038109814001550}.

\bibitem[{\citenamefont{Chen et~al.}(2020)\citenamefont{Chen, Tse, Krivorotov,
  and Lu}}]{D0NR06590K}
\bibinfo{author}{\bibfnamefont{J.-R.} \bibnamefont{Chen}},
  \bibinfo{author}{\bibfnamefont{P.~L.} \bibnamefont{Tse}},
  \bibinfo{author}{\bibfnamefont{I.~N.} \bibnamefont{Krivorotov}},
  \bibnamefont{and} \bibinfo{author}{\bibfnamefont{J.~G.} \bibnamefont{Lu}},
  \bibinfo{journal}{Nanoscale} \textbf{\bibinfo{volume}{12}},
  \bibinfo{pages}{22958} (\bibinfo{year}{2020}),
  \urlprefix\url{http://dx.doi.org/10.1039/D0NR06590K}.

\bibitem[{\citenamefont{Marra et~al.}(2016)\citenamefont{Marra, Citro, and
  Braggio}}]{PhysRevB.93.220507}
\bibinfo{author}{\bibfnamefont{P.}~\bibnamefont{Marra}},
  \bibinfo{author}{\bibfnamefont{R.}~\bibnamefont{Citro}}, \bibnamefont{and}
  \bibinfo{author}{\bibfnamefont{A.}~\bibnamefont{Braggio}},
  \bibinfo{journal}{Phys. Rev. B} \textbf{\bibinfo{volume}{93}},
  \bibinfo{pages}{220507} (\bibinfo{year}{2016}),
  \urlprefix\url{https://link.aps.org/doi/10.1103/PhysRevB.93.220507}.

\bibitem[{\citenamefont{Bena}(2017)}]{BENA2017349}
\bibinfo{author}{\bibfnamefont{C.}~\bibnamefont{Bena}},
  \bibinfo{journal}{Comptes Rendus Physique} \textbf{\bibinfo{volume}{18}},
  \bibinfo{pages}{349} (\bibinfo{year}{2017}), ISSN \bibinfo{issn}{1631-0705},
  \bibinfo{note}{2016 Prizes of the French Academy of Sciences /Prix 2016 de
  l’Académie des sciences},
  \urlprefix\url{https://www.sciencedirect.com/science/article/pii/S1631070517300579}.

\bibitem[{\citenamefont{G\l{}odzik et~al.}(2020)\citenamefont{G\l{}odzik,
  Sedlmayr, and Doma\ifmmode~\acute{n}\else
  \'{n}\fi{}ski}}]{PhysRevB.102.085411}
\bibinfo{author}{\bibfnamefont{S.}~\bibnamefont{G\l{}odzik}},
  \bibinfo{author}{\bibfnamefont{N.}~\bibnamefont{Sedlmayr}}, \bibnamefont{and}
  \bibinfo{author}{\bibfnamefont{T.}~\bibnamefont{Doma\ifmmode~\acute{n}\else
  \'{n}\fi{}ski}}, \bibinfo{journal}{Phys. Rev. B}
  \textbf{\bibinfo{volume}{102}}, \bibinfo{pages}{085411}
  (\bibinfo{year}{2020}),
  \urlprefix\url{https://link.aps.org/doi/10.1103/PhysRevB.102.085411}.

\bibitem[{\citenamefont{Maiellaro et~al.}(2021)\citenamefont{Maiellaro, Romeo,
  and Citro}}]{MaiellaroEPJplus}
\bibinfo{author}{\bibfnamefont{A.}~\bibnamefont{Maiellaro}},
  \bibinfo{author}{\bibfnamefont{F.}~\bibnamefont{Romeo}}, \bibnamefont{and}
  \bibinfo{author}{\bibfnamefont{R.}~\bibnamefont{Citro}},
  \bibinfo{journal}{Eur. Phys. J. Plus} \textbf{\bibinfo{volume}{136}},
  \bibinfo{pages}{627} (\bibinfo{year}{2021}),
  \urlprefix\url{https://doi.org/10.1140/epjp/s13360-021-01592-9}.

\bibitem[{\citenamefont{De~Luca et~al.}(2014)\citenamefont{De~Luca, Di~Capua,
  Di~Gennaro, Granozio, Stornaiuolo, Salluzzo, Gadaleta, Pallecchi, Marr\`e,
  Piamonteze et~al.}}]{deluca_2014}
\bibinfo{author}{\bibfnamefont{G.~M.} \bibnamefont{De~Luca}},
  \bibinfo{author}{\bibfnamefont{R.}~\bibnamefont{Di~Capua}},
  \bibinfo{author}{\bibfnamefont{E.}~\bibnamefont{Di~Gennaro}},
  \bibinfo{author}{\bibfnamefont{F.~M.} \bibnamefont{Granozio}},
  \bibinfo{author}{\bibfnamefont{D.}~\bibnamefont{Stornaiuolo}},
  \bibinfo{author}{\bibfnamefont{M.}~\bibnamefont{Salluzzo}},
  \bibinfo{author}{\bibfnamefont{A.}~\bibnamefont{Gadaleta}},
  \bibinfo{author}{\bibfnamefont{I.}~\bibnamefont{Pallecchi}},
  \bibinfo{author}{\bibfnamefont{D.}~\bibnamefont{Marr\`e}},
  \bibinfo{author}{\bibfnamefont{C.}~\bibnamefont{Piamonteze}},
  \bibnamefont{et~al.}, \bibinfo{journal}{Phys. Rev. B}
  \textbf{\bibinfo{volume}{89}}, \bibinfo{pages}{224413}
  (\bibinfo{year}{2014}),
  \urlprefix\url{https://link.aps.org/doi/10.1103/PhysRevB.89.224413}.

\bibitem[{Note1()}]{Note1}
Note1, \bibinfo{note}{both Refs. \cite {NatCommun14984,NPJ066} consider a
  magnetic field along the $z$-direction. We have verified that changing the
  magnetic field direction from $x$ to $z$ has no physical significance in the
  emerging phenomenon and the latter considerations still remains true.
  However, in view of the non-trivial multiband effects, small quantitative
  differences are shown for $M_i \geq 0.3$ ($i=x,z$). In Appendix \ref {AppF}
  we compare the two directions of magnetic field.}

\bibitem[{\citenamefont{Nichele et~al.}(2020)\citenamefont{Nichele, Portol\'es,
  Fornieri, Whiticar, Drachmann, Gronin, Wang, Gardner, Thomas, Hatke
  et~al.}}]{PhysRevLett.124.226801}
\bibinfo{author}{\bibfnamefont{F.}~\bibnamefont{Nichele}},
  \bibinfo{author}{\bibfnamefont{E.}~\bibnamefont{Portol\'es}},
  \bibinfo{author}{\bibfnamefont{A.}~\bibnamefont{Fornieri}},
  \bibinfo{author}{\bibfnamefont{A.~M.} \bibnamefont{Whiticar}},
  \bibinfo{author}{\bibfnamefont{A.~C.~C.} \bibnamefont{Drachmann}},
  \bibinfo{author}{\bibfnamefont{S.}~\bibnamefont{Gronin}},
  \bibinfo{author}{\bibfnamefont{T.}~\bibnamefont{Wang}},
  \bibinfo{author}{\bibfnamefont{G.~C.} \bibnamefont{Gardner}},
  \bibinfo{author}{\bibfnamefont{C.}~\bibnamefont{Thomas}},
  \bibinfo{author}{\bibfnamefont{A.~T.} \bibnamefont{Hatke}},
  \bibnamefont{et~al.}, \bibinfo{journal}{Phys. Rev. Lett.}
  \textbf{\bibinfo{volume}{124}}, \bibinfo{pages}{226801}
  (\bibinfo{year}{2020}),
  \urlprefix\url{https://link.aps.org/doi/10.1103/PhysRevLett.124.226801}.

\bibitem[{\citenamefont{Kashiwaya et~al.}(2019)\citenamefont{Kashiwaya, Saitoh,
  Kashiwaya, Koyanagi, Sato, Yada, Tanaka, and Maeno}}]{Kashiwaya19}
\bibinfo{author}{\bibfnamefont{S.}~\bibnamefont{Kashiwaya}},
  \bibinfo{author}{\bibfnamefont{K.}~\bibnamefont{Saitoh}},
  \bibinfo{author}{\bibfnamefont{H.}~\bibnamefont{Kashiwaya}},
  \bibinfo{author}{\bibfnamefont{M.}~\bibnamefont{Koyanagi}},
  \bibinfo{author}{\bibfnamefont{M.}~\bibnamefont{Sato}},
  \bibinfo{author}{\bibfnamefont{K.}~\bibnamefont{Yada}},
  \bibinfo{author}{\bibfnamefont{Y.}~\bibnamefont{Tanaka}}, \bibnamefont{and}
  \bibinfo{author}{\bibfnamefont{Y.}~\bibnamefont{Maeno}},
  \bibinfo{journal}{Phys. Rev. B} \textbf{\bibinfo{volume}{100}},
  \bibinfo{pages}{094530} (\bibinfo{year}{2019}),
  \urlprefix\url{https://link.aps.org/doi/10.1103/PhysRevB.100.094530}.

\bibitem[{\citenamefont{Golubov et~al.}(2004)\citenamefont{Golubov, Kupriyanov,
  and Il'ichev}}]{Golubov04}
\bibinfo{author}{\bibfnamefont{A.~A.} \bibnamefont{Golubov}},
  \bibinfo{author}{\bibfnamefont{M.~Y.} \bibnamefont{Kupriyanov}},
  \bibnamefont{and} \bibinfo{author}{\bibfnamefont{E.}~\bibnamefont{Il'ichev}},
  \bibinfo{journal}{Rev. Mod. Phys.} \textbf{\bibinfo{volume}{76}},
  \bibinfo{pages}{411} (\bibinfo{year}{2004}),
  \urlprefix\url{https://link.aps.org/doi/10.1103/RevModPhys.76.411}.

\bibitem[{\citenamefont{Barone and Paterno}(1982)}]{Barone82}
\bibinfo{author}{\bibfnamefont{A.}~\bibnamefont{Barone}} \bibnamefont{and}
  \bibinfo{author}{\bibfnamefont{G.}~\bibnamefont{Paterno}},
  \emph{\bibinfo{title}{Physics and applications of the Josephson effect}}
  (\bibinfo{publisher}{Wiley, New York}, \bibinfo{year}{1982}),
  \urlprefix\url{https://onlinelibrary.wiley.com/doi/abs/10.1002/352760278X}.

\end{thebibliography}

\end{document}